\documentclass[usenatbib]{mn2e}
\usepackage{natbibmnfix,graphicx,times}

\def\myputfigure#1#2#3#4#5%
{\vskip#5pt\makebox[0pt]{\hskip#2in
\includegraphics[width=#3\textwidth]{#1}}\vskip#4pt\hfill}

\newcommand\lsim{\mathrel{\rlap{\lower4pt\hbox{\hskip1pt$\sim$}}
        \raise1pt\hbox{$<$}}}
\newcommand\gsim{\mathrel{\rlap{\lower4pt\hbox{\hskip1pt$\sim$}}
        \raise1pt\hbox{$>$}}}

\newcommand{\avenf}{\bar{x}_{\rm HI}}

\newcommand{\lya}{Ly$\alpha$}

\newcommand{\taudamp}{\tau_{D}}

\newcommand{\lobs}{\lambda_{\rm obs}}

\newcommand{\zsource}{z_{\rm s}}

\newcommand{\volavenf}{\bar{x}_{\rm HI}}

\newcommand{\Msun}{M_\odot}
\newcommand{\Tvir}{T_{\rm vir}}

\newcommand{\Mmin}{M_{\rm min}}
\newcommand{\zbegin}{z_{\rm begin}}
\newcommand{\zend}{z_{\rm end}}
\newcommand{\Tgamma}{T_{\gamma, \rm res}}
\newcommand{\avexi}{\bar{\xi}_{12}}

\begin{document}


\title[Lyman-$\alpha$ Emitters During Reionization]{Lyman-$\alpha$ Emitters During the Early Stages of Reionization}

\author[Mesinger \& Furlanetto]{Andrei Mesinger\thanks{Email: andrei.mesinger@yale.edu} \& Steven R. Furlanetto \\
Yale Center for Astronomy and Astrophysics, Yale University, PO Box 208121, New Haven, CT 06520-8121}

\voffset-.6in

\maketitle

\begin{abstract}
We investigate the potential of exploiting \lya\ Emitters (LAEs) to constrain the volume-weighted mean neutral hydrogen fraction of the IGM, $\avenf$, at high redshifts (specifically $z\sim9$). We use ``semi-numerical'' simulations to efficiently generate density, velocity, and halo fields at $z=9$ in a 250 Mpc \
box, resolving halos with masses $M\ge2.2\times 10^8 \Msun$.  We construct ionization fields corresponding to various values of $\avenf$.  With these, we generate LAE luminosity functions and ``counts-in-cell'' statistics.  As in previous studies, we find that LAEs begin to disappear rapidly when $\avenf \gsim 0.5$.  Constraining $\avenf(z=9)$ with luminosity functions is difficult due to the many uncertainties inherent in the host halo mass $\leftrightarrow$ \lya\ luminosity mapping.  However, using a very conservative mapping, we show that the number densities derived using the six $z\sim9$ LAEs recently discovered by Stark et al. (2007) imply $\avenf \lsim 0.7$.  On a more fundamental level, these LAE number densities, if genuine, \emph{require} substantial star formation in halos with $M \lsim 10^9 \Msun$, making them unique among the current sample of observed high-$z$ objects. Furthermore, reionization increases the apparent clustering of the observed LAEs.  We show that a ``counts-in-cell'' statistic is a powerful probe of this effect, especially in the early stages of reionization.  Specifically, we show that a field of view (typical of upcoming IR instruments) containing LAEs has $\gsim$10\% higher probability of containing {\it more than one} LAE in a $\avenf\gsim0.5$ universe than a $\avenf \approx 0$ universe with the same overall number density.  
With this statistic, an ionized universe can be robustly distinguished from one with $\avenf \ga 0.5$ using a survey containing only $\sim$ 20--100 galaxies.
\end{abstract}


\section{Introduction}
\label{sec:intro}

The reionization of hydrogen in the intergalactic medium (IGM) is a landmark event in the early history of structure formation, because it defines the moment at which galaxies (and black holes) affected every baryon in the Universe.  As such, it has received a great deal of attention -- both observationally and theoretically -- in the past several years.  Unfortunately, the existing observational evidence is enigmatic (see \citealt{fan06-review} for a recent review).  
Electron scattering of cosmic microwave background photons implies that reionization occurred at $z \sim 10$, albeit with a large uncertainty \citep{Page06}.  On the other hand, quasars at $z \sim 6$ show some evidence for a rapid transition in the globally-averaged neutral fraction, $\volavenf$ (e.g., \citealt{Fan06, MH04, MH07}).  However the \lya\ absorption is so saturated in the Gunn-Peterson (GP) trough that constraints derived from that spectral region \citep{Fan06, Maselli07} are difficult to interpret (e.g, \citealt{lidz06, becker07, bolton07}).  

Of particular recent interest have been efforts to constrain reionization (and star formation at high redshifts) through searches for distant galaxies.  Currently, the most efficient way to find distant galaxies is by searching for \lya\ emission lines (which result from the re-processing of ionizing photons inside the galaxy; \citealt{partridge67}); such surveys now routinely reach $z \gsim 6$, where constraints on reionization become interesting (e.g., \citealt{hu02-lya, Kodaira03, rhoads04,santos04-obs,stanway04, taniguchi05, Kashikawa06}), and are now being stretched to even higher redshifts \citep{willis05, iye06, cuby07, Stark07, ota07}.  They offer a number of advantages over more traditional techniques.  First, narrowband searches reduce the sky background, especially if placed between the bright sky lines that (nearly) blanket the near-infrared sky (e.g., \citealt{barton04}).  Second, they efficiently select galaxies
at a known redshift (albeit with some contamination by lower-redshift interlopers).  Third, they increase the signal-to-noise by focusing on an emission line.  Of course, the disadvantage is that extraordinarily deep spectroscopic follow-up is required to study the detailed properties of the sources (for an illustration of the difficulties, see \citealt{Stark07}). 

However, these properties are relatively unimportant for studying the ionization state of the IGM, which can be measured from the \lya\ line photons themselves.  In particular, these are absorbed if they pass through neutral gas near the galaxy.  This is a consequence of the enormous \lya\ optical depth of a neutral IGM: $\tau_{\rm GP} \sim 6.5 \times 10^5 \, \volavenf \, [(1+z)/10]^{3/2}$ \citep{GP65}, so even those photons passing through the damping wing of the \lya\ resonance will be absorbed \citep{miralda98}.

Thus, as the IGM becomes more neutral, the \lya\ selection technique will detect fewer and fewer objects (even after accounting for cosmological evolution in their intrinsic abundance); the number of such galaxies therefore measures $\volavenf$ \citep{HS99}.
  The optical depth encountered by a galaxy's \lya\ photons depends primarily on the extent of the HII region that surrounds it: the photons redshift as they stream through the ionized gas (suffering little absorption), so they are somewhere in the wings of the line by the time they encounter the neutral gas.  Thus, the amount of absorption depends sensitively on the size distribution of ionized bubbles during reionization.  Early work treated each galaxy or quasar in isolation \citep{madau00,Haiman02,santos04,HC05}, so the HII regions were rather small even late in reionization.  Including clustering dramatically increases the sizes of the ionized bubbles \citep{FZH04}, which allows \lya\ galaxies to be visible farther back in reionization \citep{FHZ04, FZH06, McQuinn07LAE}.

The current observational picture is ambiguous.  \citet{MR04} compared luminosity functions of Lyman $\alpha$-emitters (LAEs) at $z=5.7$ and $z=6.5$ (bracketing the time at which quasar spectra indicate that reionization ends) and found no evolution in the number density over this range, requiring a substantial ionized fraction at $z \sim 6.5$ (see also \citealt{HC05, FZH06, MR06}).  However, \citet{Kashikawa06} found a decline in the number density of bright LAEs over the same redshift range (with a larger sample of objects), and \citet{ota07} suggest a farther decline to $z \sim 7$.  
On the other hand, \citet{dawson07} argued that the LAE luminosity function remains nearly constant from $z=3$--$7$ using a compilation of other surveys.  Moreover, \citet{Stark07} found a surprisingly high abundance of faint emitters at $z \sim 9$.  These objects are of particular interest to us, because they present the best possibility of being affected by reionization.  

The interpretation of these results is complicated by the unknown internal properties of the sources:  in particular, how do their \lya\ line luminosities depend on the underlying halo mass, and what other factors affect the mapping?  A number of recent studies have developed models for the high-redshift LAE population \citep{ledelliou06, DWH06, fernandez07, SLE07, kobayashi07}, but the relevant parameters (such as the star formation efficiency and time-scale and the initial mass function) are still mostly unconstrained, especially if they evolve.  Indeed, \citet{DWH06} have argued that the declining number density of bright LAEs observed by \citet{Kashikawa06} can be simply a result of the evolving mass function.  On the other hand, \citet{Kashikawa06} argued that the LAE luminosity function evolved more than that of Lyman-break selected galaxies over this range, so that the decline could not be attributed to the galaxies themselves (see also \citealt{ota07}).  
Most recently, \citet{dawson07} argued that the LAEs evolve less rapidly than other galaxy samples.  In this paper, we will present new simulations of the luminosity function evolution throughout reionization, focusing on the high-redshift ``frontier" at $z=9$, and try to place some \emph{model-independent} constraints on the sources and the IGM.

However, because of the difficulties  involved, it is advantageous to consider other signatures of reionization.  In particular, the clustering of LAEs can be a powerful probe of reionization \citep{FHZ04, FZH06, McQuinn07LAE}.  This is because their visibility is modulated by the pattern of ionized bubbles:  large bubbles (which surround overdensities with many galaxies) have a small damping-wing optical depth, so that most sources will be visible, while small bubbles (where galaxies are rare to begin with) will appear to be entirely empty.  Thus, during reionization observable LAEs should appear \emph{more} clustered than the underlying population, with the boost decreasing as the ionized bubbles grow (which happens relatively quickly).  Clustering of the overall galaxy population should evolve much more slowly than number densities, so a rapid change in the clustering is a good indicator of reionization \citep{McQuinn07LAE}.

Existing work has examined clustering primarily in the context of the power spectrum (or correlation function) of the galaxies (\citealt{FZH06, McQuinn07LAE}; see also the appendix in \citealt{McQuinn07LAE} for a brief discussion on void and peak statistics).
Most recently, \citet{McQuinn07LAE} suggest that the lack of clustering evolution in the $z=5.7$ and $z=6.6$ LAEs rules out a substantially neutral universe at $z=6.6$.  However, the bubble modulation is actually highly non-gaussian.  It is therefore useful to consider other statistics, especially early in reionization (when sources are so rare that measuring the scale-dependent power spectrum robustly will be extremely difficult).  Here we take a first step in this direction by considering a ``counts-in-cells" measurement of the clustering.

This paper is organized as follows.  In \S \ref{sec:sim}, we briefly outline the main points of our ``semi-numerical'' simulation, originally presented in \citet{MF07}. In \S \ref{sec:taud} we present the \lya\ optical depth distributions of galaxies in our simulations, as functions of halo mass and $\avenf$.  In \S \ref{sec:lf} we show the resulting LAE luminosity functions, comparing with the \citet{Stark07} sample in \S \ref{sec:obs}.  In \S \ref{sec:incell}, we present statistics using ``counts in-cell''.  Finally in \S \ref{sec:conc} we summarize our key findings and offer some conclusions.

Unless stated otherwise, we quote all quantities in comoving units.
We adopt the background cosmological parameters
($\Omega_\Lambda$, $\Omega_{\rm M}$, $\Omega_b$, $n$, $\sigma_8$, $H_0$)
= (0.76, 0.24, 0.0407, 1, 0.76, 72 km s$^{-1}$ Mpc$^{-1}$), consistent
with the three--year results of the {\it WMAP} satellite
\citep{Spergel06}.

\section{Semi-Numerical Simulations}
\label{sec:sim}

We use an excursion-set approach combined with first-order Lagrangian perturbation theory to efficiently generate density, velocity, halo, and ionization fields at $z=9$.  This ``semi-numerical'' simulation is presented in \citet{MF07}, to which we refer the reader for details.
A similar halo-finding scheme has also been presented by \citet{BM96_algo} and a similar scheme to generate ionization fields has been presented by \citet{Zahn07}.

Our simulation box is 250 Mpc on a side, with the final density, velocity and ionization fields having grid cell sizes of 0.5 Mpc.  Halos with a total mass $M \ge 2.2 \times 10^8$ $\Msun$ are filtered out of the linear density field using excursion-set theory, with mass scales spaced as $\Delta M / M = 1.2$.
Note that we are able to resolve halos with masses less than a factor of two from the cooling mass likely to be pertinent mid-reionization 
(or more precisely, during the redshift interval $6\lsim z \lsim10$),
 corresponding to gas with a temperature of $T\sim10^4$ K (e.g. \citealt{Efstathiou92, TW96, Gnedin00b, SGB94}).
  Halo locations are then adjusted using first-order Lagrangian perturbation theory.
The resulting halo field matches both the mass function and statistical clustering properties of halos in N-body simulations \citep{MF07}.

\begin{figure*}
\vspace{+0\baselineskip}
{
\includegraphics[width=0.245\textwidth]{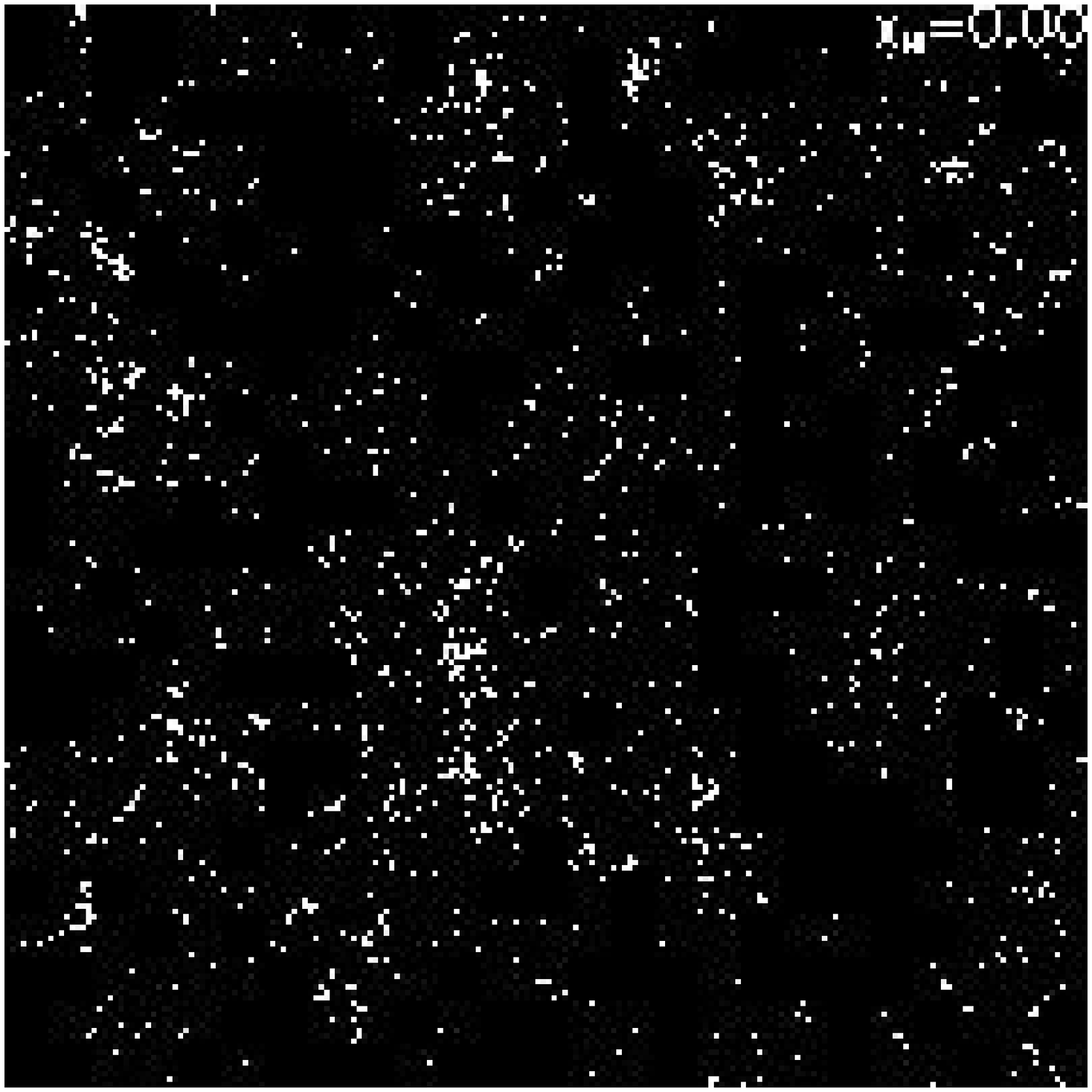}
\includegraphics[width=0.245\textwidth]{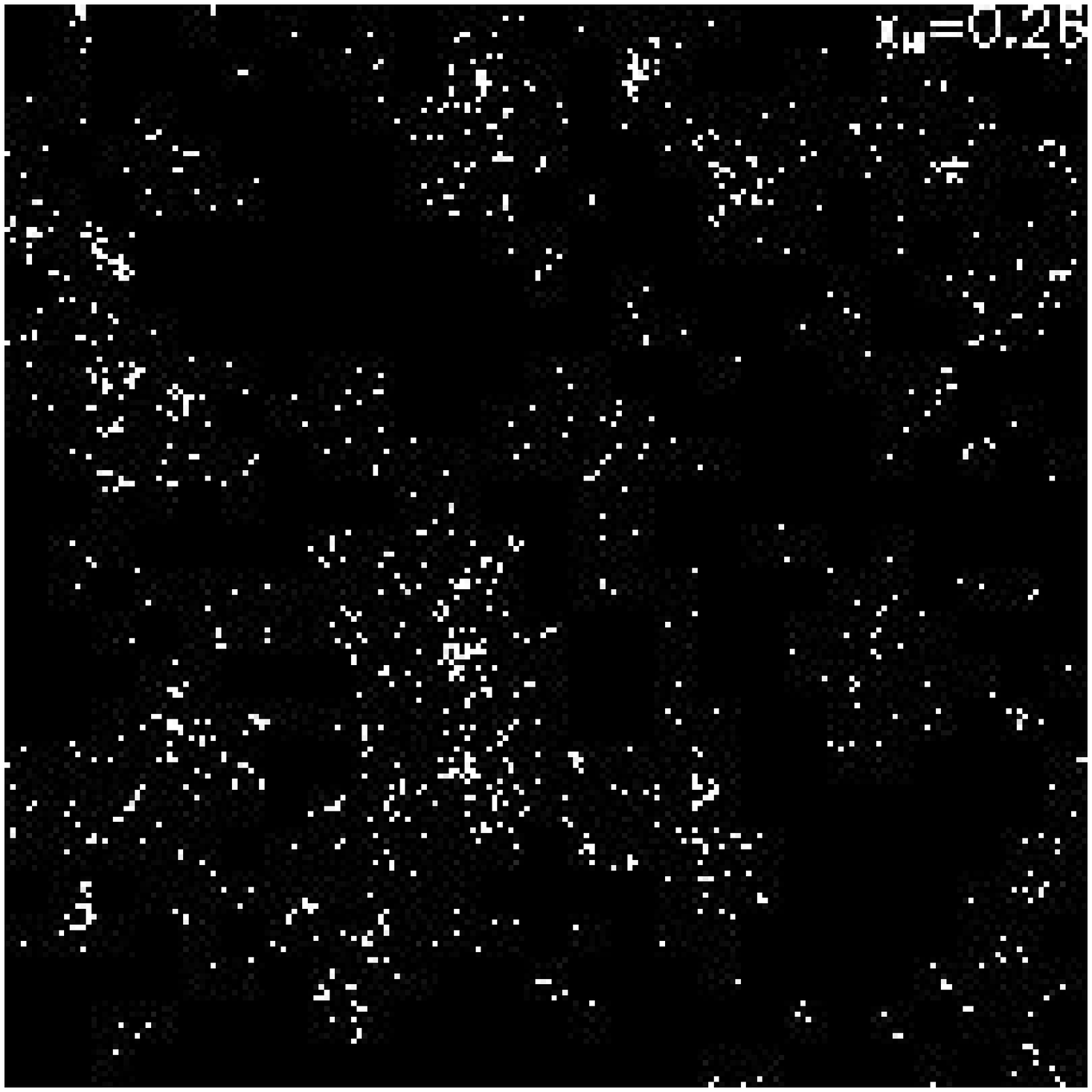}
\includegraphics[width=0.245\textwidth]{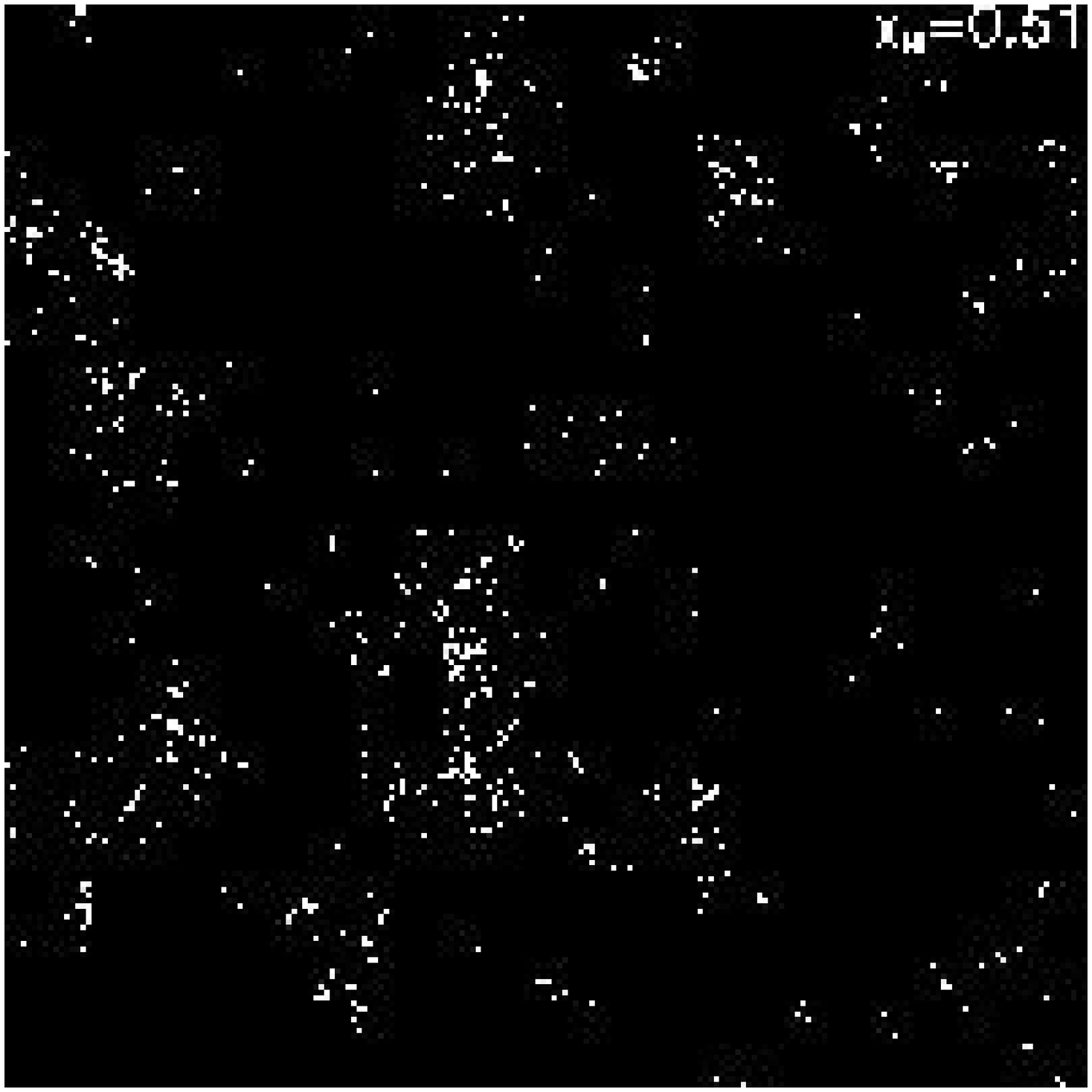}
\includegraphics[width=0.245\textwidth]{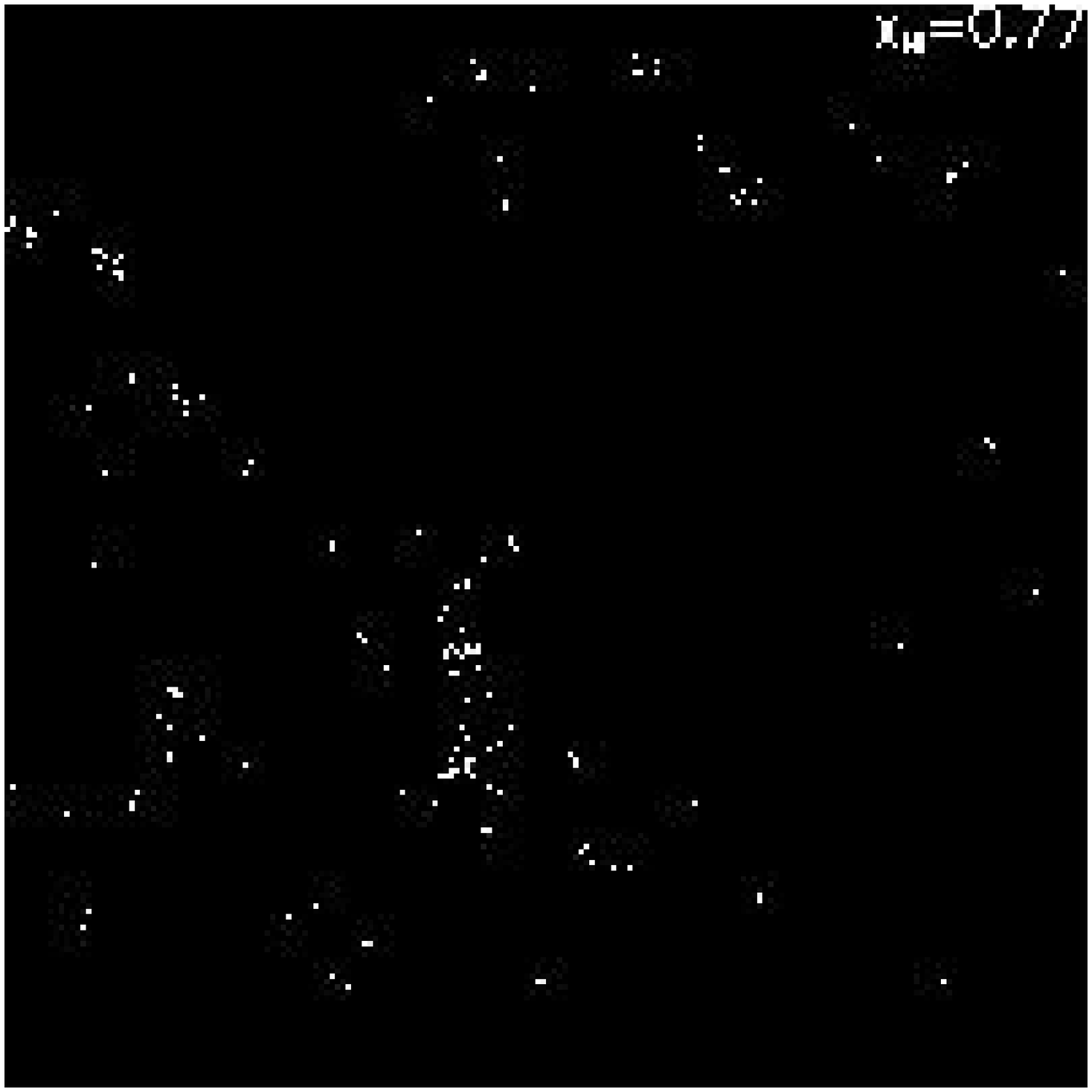}
}
\caption{
Maps of visible halos at $z=9$, assuming $\Mmin= 1.67\times10^{10}$ $\Msun$, and $\avenf \approx$ 0, 0.26, 0.51, 0.77, ({\it left to right}).  All slices are 250 Mpc on a side and 20 Mpc deep (corresponding to a narrow band filter with $R=\lambda/\Delta \lambda \sim 125$).  The 1500$^3$ halo field is smoothed onto a 200$^3$ grid here for viewing purposes.
\label{fig:maps}
}
\vspace{-1\baselineskip}
\end{figure*}

In constructing the ionization field, the IGM is modeled as a two-phase medium, comprised of fully ionized and fully neutral regions (this is a fairly accurate assumption at high-redshifts preceding the end of reionization, unless the X-ray background is rather strong).  Using the same halo field at $z=9$, we generate ionization fields corresponding to different values of $\avenf$ by varying a single efficiency parameter,\footnote{This differs from the method recently used by \citet{McQuinn07LAE}, who used a suite of N-body simulations with radiative transfer.  They also argued that a faster but effective method was to generate ionization fields at several different redshifts (using a single radiative transfer simulation) but apply them to a halo field at a single redshift.  They thus assumed that the ionization topology is only a weak function of redshift \citep{McQuinn07}.  The speed of our approach, which does not require a radiative transfer algorithm, allows us to generate ionization fields at a {\it single} redshift self-consistently, using the same halo field, merely by adjusting the source efficiencies.  However, we confirm that the ionization maps are very  nearly redshift-independent for most purposes (including those studied by \citealt{McQuinn07LAE}). The exceptions to this are the rare events occurring in $\lsim10^{-3}$ of the typical fields of view discussed in \S \ref{sec:incell}.} $\zeta$, again using the excursion-set approach (c.f. \citealt{MF07, FHZ04}).  

This semi-numeric approach is thus ideally suited to the LAE problem, because we are able to ``resolve" relatively small halos and simultaneously sample a large, representative volume of ionized bubbles.
Note that our ``simulations" do \emph{not} make any predictions (and only weak assumptions) about the \lya\ luminosities of these sources; we will discuss the mapping from halo mass (the fundamental quantity for our simulations) to observable properties below.  
This mapping must also be prescribed in state-of-the-art cosmological simulations, which cannot self-consistently include hydrodynamics (and hence star formation) while also subtending a representative volume during reionization (c.f., \citealt{McQuinn07LAE}).

\section{Damping Wing Optical Depth Distributions}
\label{sec:taud}

To study the effects of reionization, we first need to track the absorption of line photons from neutral gas in the IGM.  We divide the absorption into two parts:  the resonant and damping wing components.  This is convenient because they correspond to two spatially distinct sets of absorbers.  Resonant absorption occurs whenever a photon that begins blueward of line center redshifts into resonance (either inside the HII region surrounding the source or in the neutral gas outside).  Because the line-center optical depth is so large, this component can lead to nearly complete absorption -- but only for photons on the blue side of the line (e.g., \citealt{santos04}).  We do not model this component in detail in this work, assuming that a constant fraction of the Ly$\alpha$ line gets resonantly absorbed.

On the other hand, photons that begin redward of line center only redshift farther away.  It is therefore only the damping wings of the line that affect them, and the amount of absorption, exp[-$\taudamp$], where $\taudamp$ is the damping-wing optical depth, will depend sensitively on the size of the host HII region
but be \emph{insensitive} to the precise $x_{\rm HI}$ inside the ionized region.  It is this component that evolves most rapidly through reionization.  Figure \ref{fig:maps} shows the visible halos at $z=9$, with $M$ exp[$-\tau_D] > 1.67\times10^{10} \Msun$, and $\avenf \approx$ 0, 0.26, 0.51, 0.77, ({\it left to right}); the obscuration from damping wing absorption is obvious.

\begin{figure*}
\vspace{+0\baselineskip}
{
\includegraphics[width=0.245\textwidth]{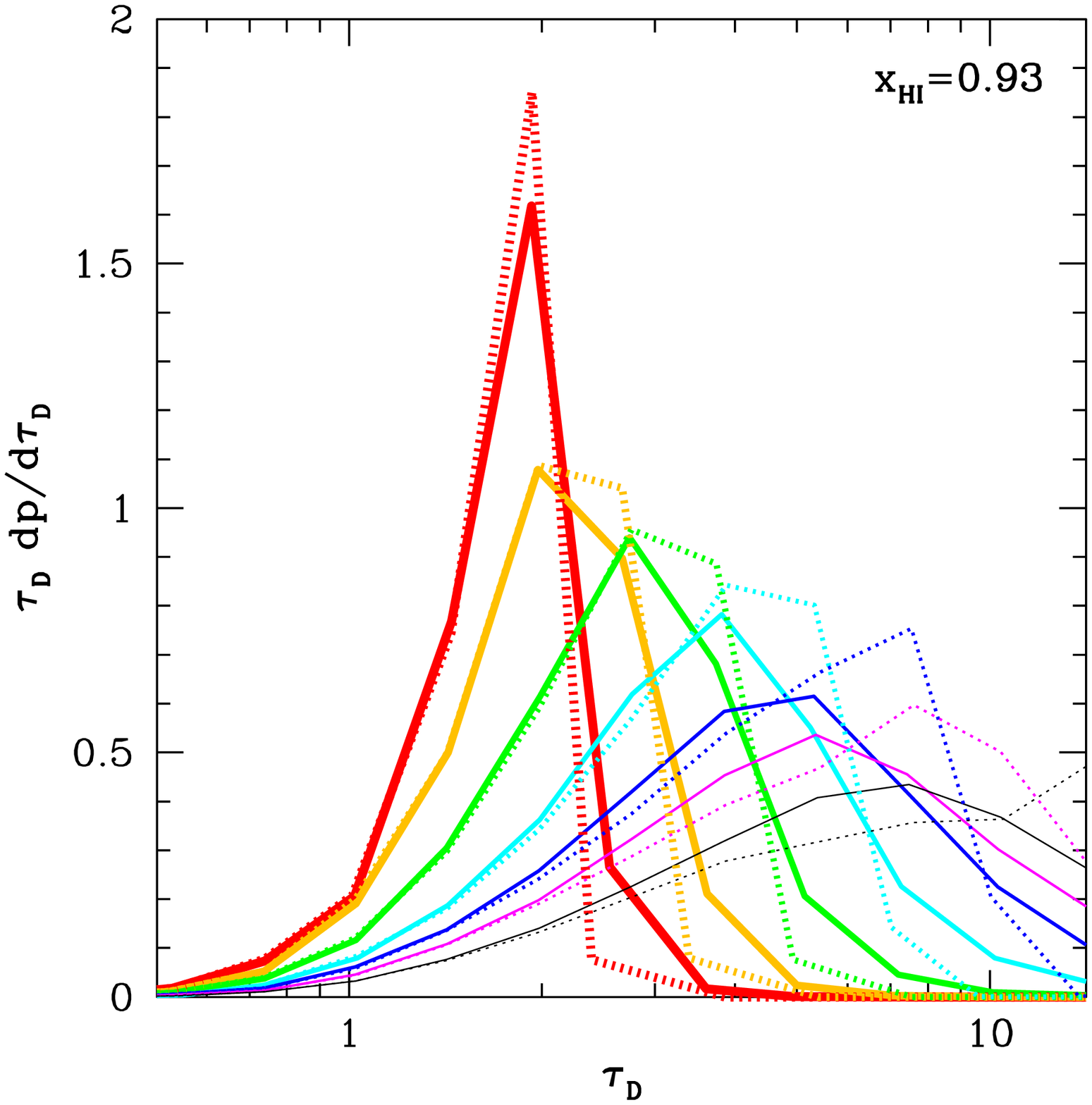}
\includegraphics[width=0.245\textwidth]{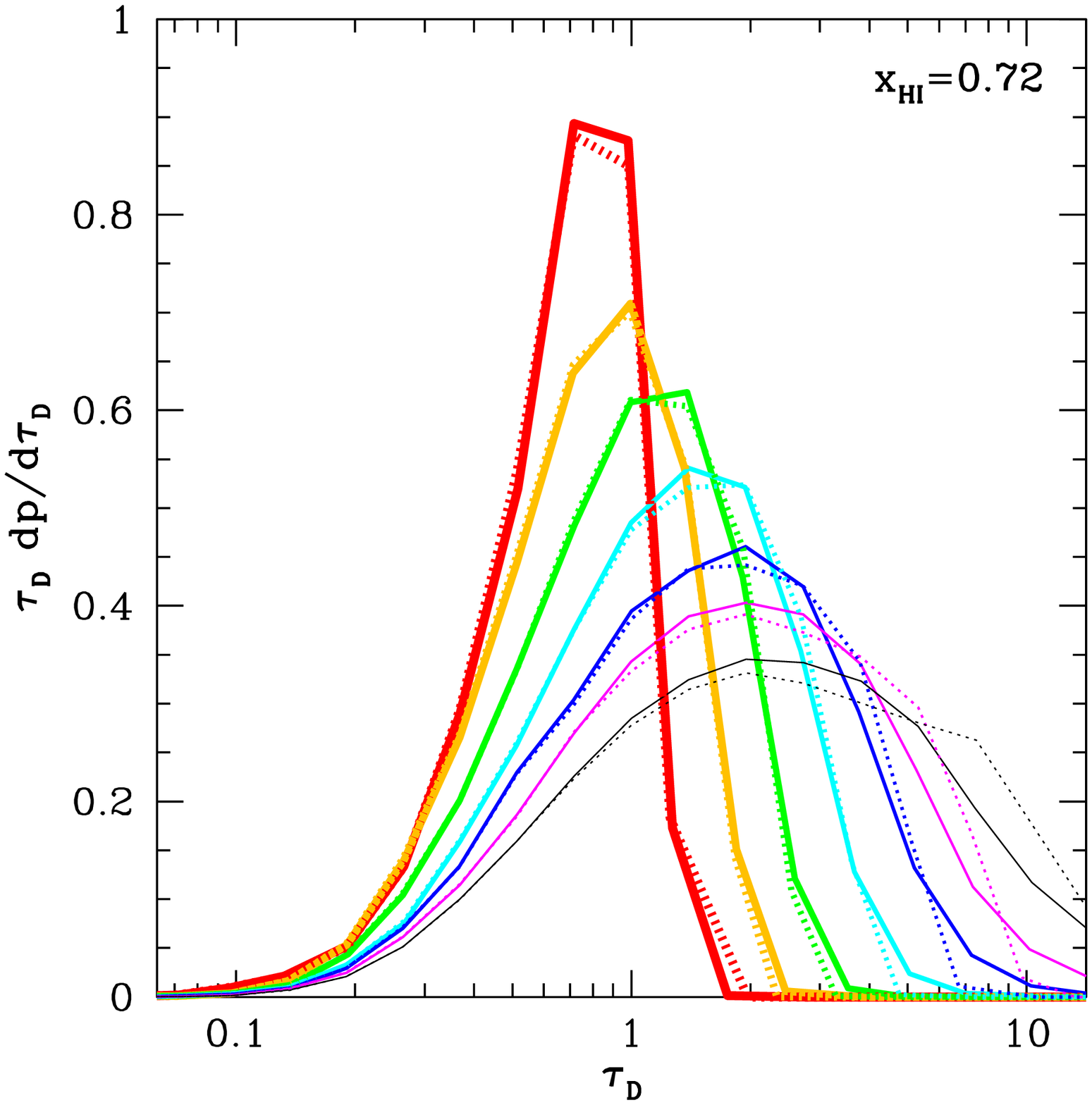}
\includegraphics[width=0.245\textwidth]{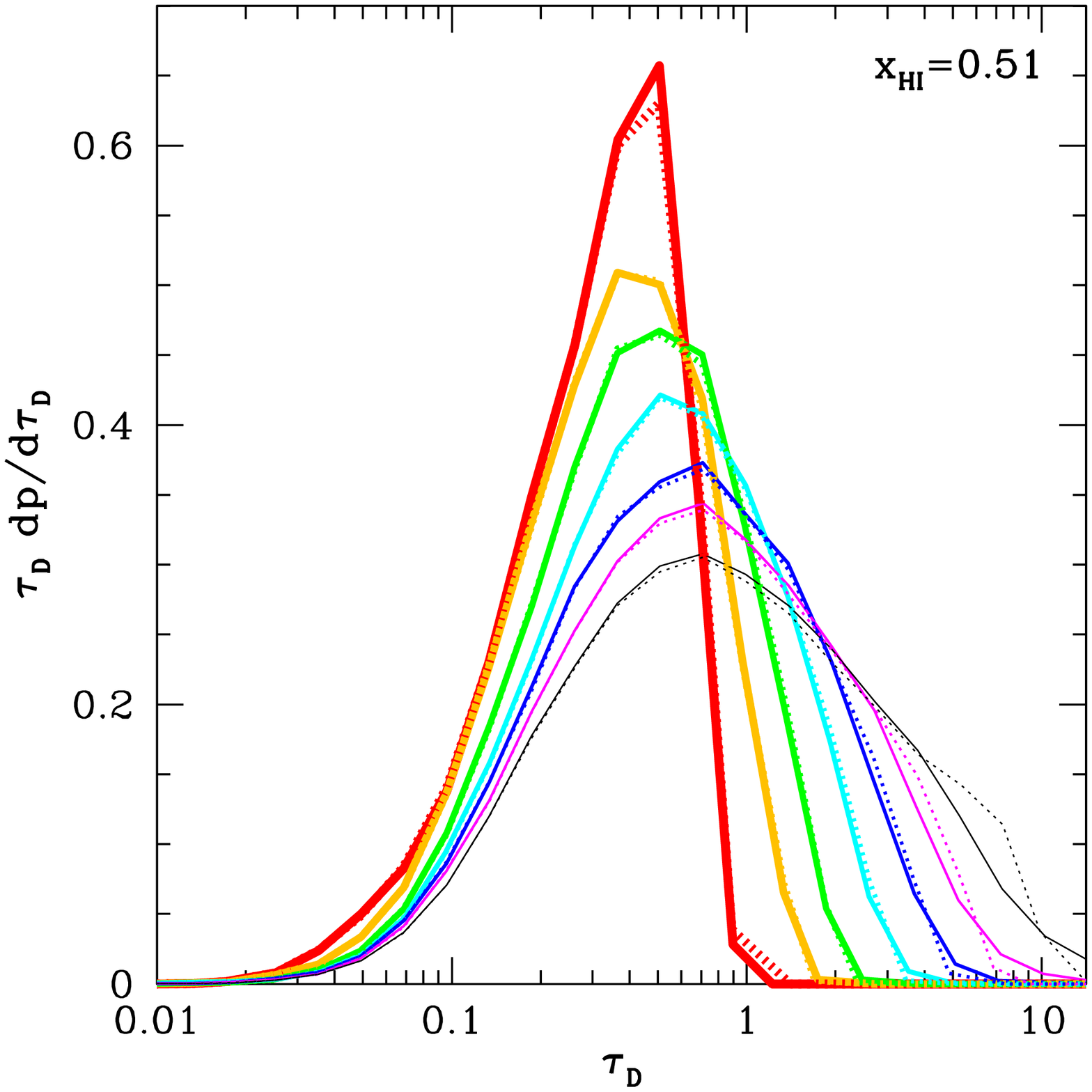}
\includegraphics[width=0.245\textwidth]{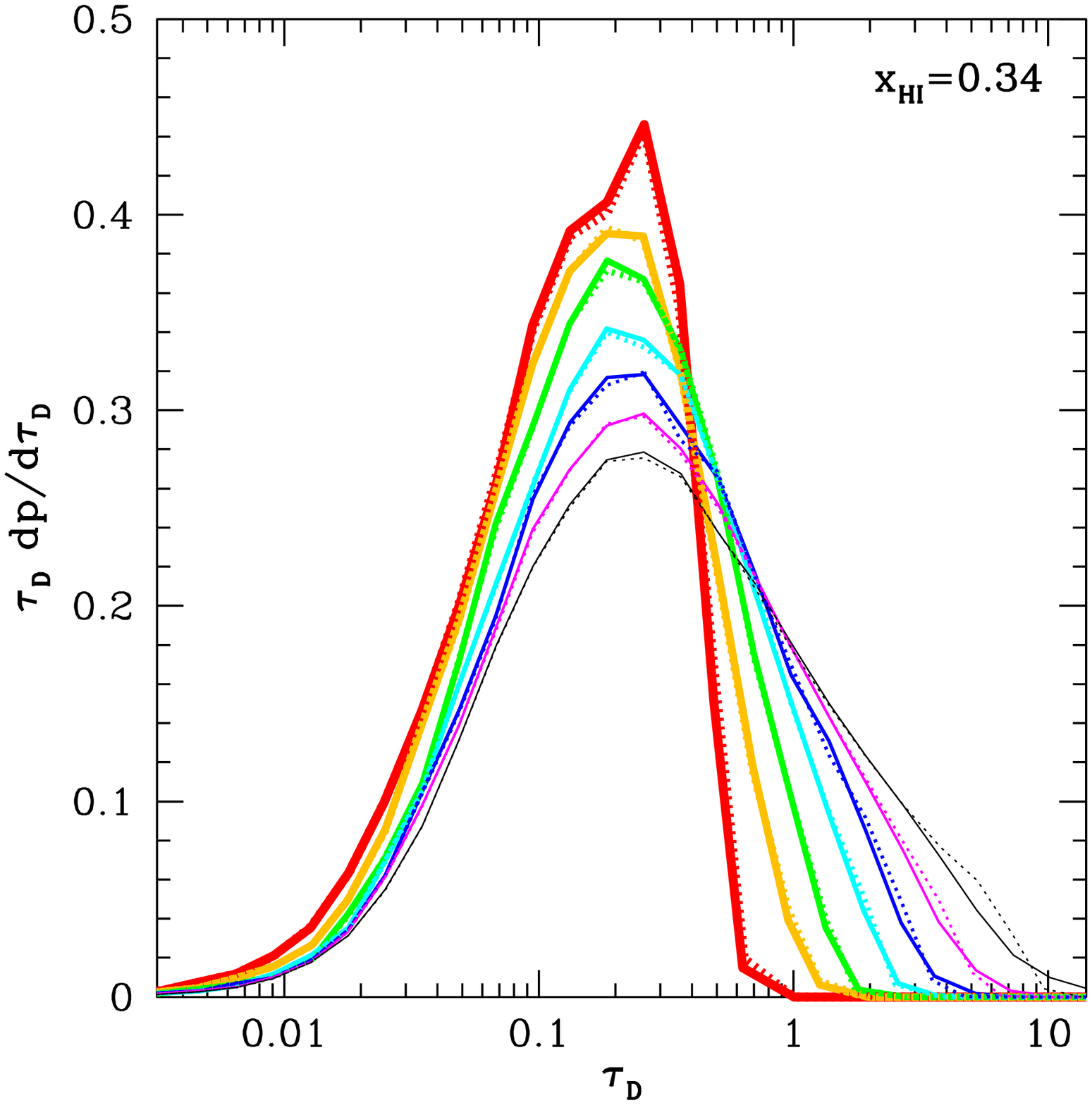}
}
\vspace{-1\baselineskip} \caption{
Damping wing optical depth distributions at $\avenf(z=9) = 0.93, 0.72, 0.51, 0.34$ ({\it left to right}).  Curves correspond to LOSs originating from halos with masses $M$ = $2.6\times10^{11}$, $1.2\times10^{11}$, $5.4\times10^{10}$, $2.5\times10^{10}$, $1.1\times10^{10}$, $5.1\times10^{9}$, and $2.3\times10^{9}$ $\Msun$ ({\it left to right}).  Solid curves include the peculiar velocity offset of the host halo; dotted curves do not.
\label{fig:taud}
}
\vspace{-1\baselineskip}
\end{figure*}

We compute the total line center
 \lya\ optical depth along a randomly chosen line-of-sight (LOS) centered on a halo location at $\zsource = 9.0$.  We do this by summing the damping wing optical depth, $\taudamp$, contribution from each neutral hydrogen patch (extending from $\zbegin$ to $\zend$) encountered along the LOS, using the approximation \citep{Miralda-Escude98}:
\begin{eqnarray}
\label{eq:jordi_tau}
\taudamp &=& 6.43 \times 10^{-9} \left( \frac{\pi e^2 f_{\alpha}n_H(\zsource)}{m_e c H(\zsource)} \right)\\
\nonumber &\times& \left[ I\left( \frac{1+\zbegin}{1+\zsource}\right) - I\left( \frac{1+\zend}{1+\zsource}\right)  \right]
\end{eqnarray}
\noindent where $n_H(\zsource)$ is the mean hydrogen number density of the IGM at redshift $\zsource$, and
\begin{eqnarray}
\nonumber I(x) &\equiv& \frac{x^{9/2}}{1-x} + \frac{9}{7} x^{7/2} + \frac{9}{5} x^{5/2} + 3x^{3/2} + 9x^{1/2} \\
\nonumber &-& \ln \left| \frac{1+x^{1/2}}{1-x^{1/2}} \right| ~ .
\end{eqnarray}
\noindent
We use eq. (\ref{eq:jordi_tau}) to calculate the optical depth for each neutral hydrogen patch, summing the contributions of patches along the LOS for 200 Mpc,\footnote{This number was chosen experimentally in order to ensure convergence of the $\taudamp$ distributions at the mass scales and neutral fractions studied in this work.} wrapping around the simulation box if needed.
  We construct distributions of $\taudamp$ for each halo mass scale and ionization topology (i.e. $\avenf$).  We make sure to process LOSs from every halo of a particular mass scale, cycling through the halo list until each mass scale undergoes a minimum of $3\times10^4$ such Monte-Carlo realizations.  We also include the component of the source halo's peculiar velocity along the LOS, $v$, in our estimates of $\taudamp$ by substituting $\zsource \rightarrow \zsource+v/c$.
\footnote{Note that for simplicity we do \emph{not} include the peculiar velocity of the neutral IGM patches, which could be correlated on large scales with the peculiar velocities of the sources, thus diminishing the impact of velocities on $\taudamp$.  However, note from Fig. \ref{fig:taud} that peculiar velocities do not play a major role in the optical depth distributions except when $\tau_D \gg 1$ and $\avenf \sim 1$ (where the absorption is so strong that its precise value does not matter for our purposes).
The treatment of velocities is uncertain in any case because we ignore the possibility of galactic winds, which can move the \lya\ lines redward and decrease the absorption \citep{santos04}.}

As delineated above, we only compute $\taudamp$ at a single observed wavelength, $\lobs = 1215.67 (1+\zsource+v/c)$ \AA, instead of over the entire wavelength extent of the \lya\ emission line.  
To zeroth order, the damping wing optical depth varies only weakly across the typical scale of the \lya \ emission line, although its shape does provide useful information about the IGM properties; we refer the interested reader to \citet{Mesinger07-grb,McQuinn07-grb}.  
We also emphasize that, in our two-phase IGM approximation, we do not model the resonance contribution to the total optical depth.  However, any {\it mass-independent} attenuation of the \lya\ line can be swept into the assumed halo mass $\leftrightarrow$ \lya\ luminosity mapping (c.f. \S \ref{sec:obs}).  Most studies (including our own in \S \ref{sec:lf}) assume such a mass-independent resonant attenuation, justifying it with the narrow halo mass range ($\sim 1$ dex) probed by present instruments/surveys (e.g. \citealt{DWH06, SLE07, McQuinn07LAE}).  We defer a detailed study of this assumption to a future work.

In Figure \ref{fig:taud}, we show some of the optical depth distributions generated by the above procedure, for $\avenf(z=9) = 0.93, 0.72, 0.51, 0.34$ ({\it left to right}).  The panels show $\taudamp \partial p(>\taudamp, M, \avenf)/\partial \taudamp$, i.e. the probability per $\ln \taudamp$ that a LOS originating from a halo of mass $M$ embedded in an IGM with a neutral fraction of $\avenf$ has an optical depth of $\taudamp$.  Curves correspond to LOSs originating from halos with masses $M$ = $2.6\times10^{11}$, $1.2\times10^{11}$, $5.4\times10^{10}$, $2.5\times10^{10}$, $1.1\times10^{10}$, $5.1\times10^{9}$, and $2.3\times10^{9}$ $\Msun$ ({\it left to right}).  Solid curves include the peculiar velocity offset of the host halo; dotted curves do not.

In general, the optical depth distributions become narrower and their mean values decrease as the halo mass increases.  This is to be expected because the halo bias is a function of mass, with the more massive halos more likely to sit in larger overdense regions -- which in turn contain more ionizing sources and larger HII bubbles than less massive halos (e.g. \citealt{FZH06, McQuinn07LAE}).  Furthermore, since the overlap of several HII bubbles makes a smaller {\it fractional} change to the size of a large bubble hosting the most massive halos, the distributions of $\taudamp$ are narrower for massive halos.  Both of these effects diminish as reionization progresses, and the $\taudamp$ distributions start merging together.

It is also evident from Fig. \ref{fig:taud} that galaxy peculiar velocities broaden the optical depth distributions, shifting their means to smaller values.  This effect is the strongest for the smallest mass halos, because they are typically surrounded by the smallest HII bubbles (whose edges are at the smallest velocity offsets from the \lya\ line center).  The effects of source halo velocities on $\taudamp$ also diminishes with decreasing $\avenf$; for $\avenf \lsim 0.7$ peculiar velocities play a negligible role in determining $\taudamp$ for halos with $M \gsim 10^9 \Msun$. 
More importantly, Fig. \ref{fig:taud} shows that peculiar velocities have virtually no impact on the optical depth distributions in the pertinent, $\tau_D \sim 1$ regime.  Hence, they should not have a noticeable effect on any of our conclusions.

\begin{figure}
\vspace{+0\baselineskip}
\myputfigure{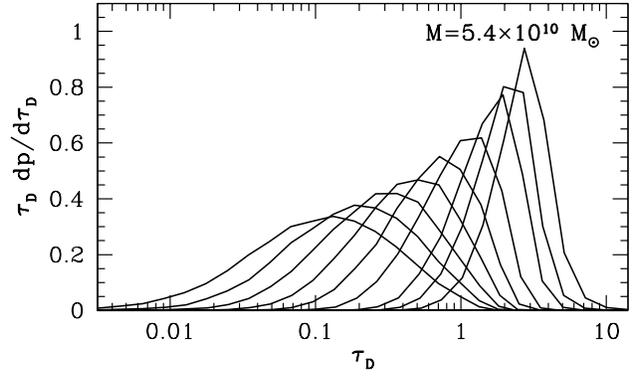}{3.3}{0.5}{.}{0.}  \caption{ 
Damping wing optical depth distributions generated using LOSs originating from halos with masses $M$ = $5.4\times10^{10} \Msun$.  Curves correspond to $\avenf = 0.26, 0.34, 0.42, 0.51, 0.61, 0.72, 0.83, 0.88, 0.93$, left to right.
\label{fig:taudvsnf}}
\vspace{-1\baselineskip}
\end{figure}

Figure \ref{fig:taudvsnf} quantifies the evolution of the optical depth distributions with $\avenf$.  The distributions were generated using source halos at a fixed mass scale of $M$ = $5.4\times10^{10} \Msun$.  Curves correspond to $\avenf = 0.26, 0.34, 0.42, 0.51, 0.61, 0.72, 0.83, 0.88, 0.93$, from left to right.  The curves illustrate that the more complex topologies and large scatter among disparate LOSs at lower values of $\avenf$ significantly broaden the distributions of $\taudamp$, in addition to decreasing their means.  This is not because the bubble size distribution widens (in fact, it narrows as reionization progresses; \citealt{FMH05}) but because some LOSs pass almost entirely through neighboring ionized bubbles, while others pass through long skewers of neutral gas.  A similar effect has been shown to have important consequences for the interpretation of quasar spectra \citep{lidz07}.

\subsection{Comparison to an Analytic Model}
\label{sec:comp-analytic}

\citet{FHZ04} presented similar optical depth distributions calculated via the analytic \citet{FZH04} model for the ionized bubbles during reionization.  Our ``semi-numeric" simulation is also based on this model, but it includes additional effects such as the complicated, non-spherical geometry of the HII regions.  It is therefore illuminating to compare our distributions with those of the purely analytic model.  For the latter, we follow the calculation of \citet{FHZ04} except that for simplicity we assume that all ionizing sources sit in the center of their bubble.  The damping wing optical depth distributions then follow from arguments similar to the ``extended Press-Schechter" formalism \citep{PS74, Bond91, LC93}.

\begin{figure}
\vspace{+0\baselineskip}
\myputfigure{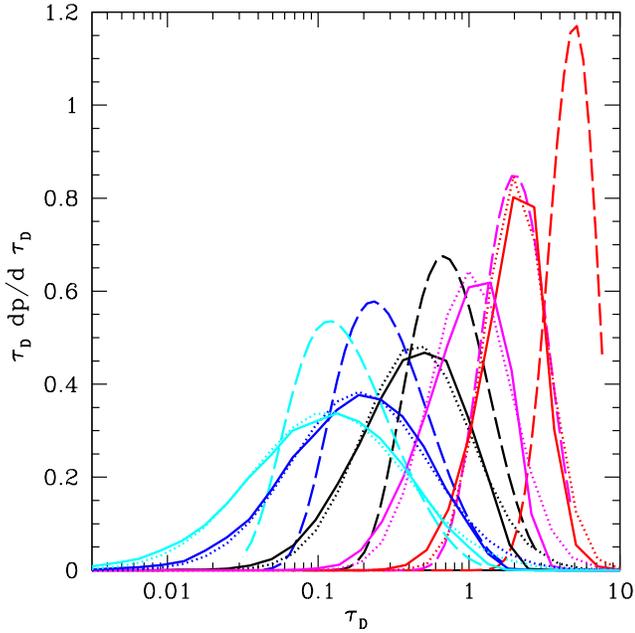}{3.3}{0.5}{.}{0.}  \caption{ 
Damping wing optical depth distributions for halos with masses $M$ = $5.4\times10^{10} \Msun$. The solid curves are computed from our simulations, and the dotted curves are the corresponding best-fit lognormal distributions.  The dashed curves show the predictions of the analytic model.  Within each set, the curves correspond to $\avenf = 0.26, 0.34, 0.51, 0.72, 0.88$, from left to right.
\label{fig:taufits}}
\vspace{-1\baselineskip}
\end{figure}

Figure~\ref{fig:taufits} shows our results for halos with $M$ = $5.4\times10^{10} \Msun$ and a variety of neutral fractions.  The solid curves are again taken from the simulations.  The dashed curves show the predictions of the analytic model.  There are two key differences with the simulated results.  First, the analytic model predicts that the distribution peaks at larger $\tau_D$ than the simulation.  This is not a surprise:  \citet{MF07} showed that, when defined in this way, the simulations have larger ionized bubbles early in reionization.  This is because of neighboring bubbles that slightly overlap; the spherical geometry required by the analytic model does not effectively capture such events.  This can reduce the apparent optical depth by a factor of $\sim 3$ early on, so it is not a small effect.  

Second, the analytic model also under-predicts the width of the distribution, particularly on the small-$\tau_D$ tail at the end of reionization.  This is significantly larger than the differences in the distributions of bubble sizes.  Instead, it likely results from the analytic model's assumption of a uniformly ionized medium outside of the source's host bubble:  as described above, many lines of sight will pass through nearby, fully-ionized bubbles, allowing for the existence of many (nearly) clear lines of sight
(see \citealt{Mesinger07-grb,McQuinn07-grb}).

Because the analytic model does not provide a good fit, we searched for a better representation of the distributions.  We find that our results are nearly always well-fit by log-normal distributions,
\begin{equation}
{\partial p(>\tau_D) \over \partial \tau_D} = {1 \over \tau_D \sqrt{ 2 \pi \sigma_D^2}} \exp \left[ - \frac{(\ln \tau_D - \mu_D)^2}{2 \sigma_D^2} \right],
\label{eq:lndist}
\end{equation}
where $\mu_D$ and $\sigma_D$ are parameters determined by fits to the simulated distributions.  Some example fits are shown by the dotted curves in Figure~\ref{fig:taufits}; in general, the lognormal distribution overestimates the strength of the tail at large $\tau_D$ and underestimates its strength at small $\tau_D$, with the skewness increasing for smaller halos and larger neutral fractions.  The following simple bilinear fits for the parameters are accurate to $\lsim 10\%$
 over the entire mass range of our simulations and from $\volavenf = 0.26$--$0.93$:
\begin{eqnarray}
\mu_D & = & -3.37 + \log M_{10} (-0.115 - 0.587 \volavenf) + 5.30 \volavenf 
\label{eq:mufit} \\
\sigma_D & = & 1.68 + \log M_{10} (-0.155 - 0.265 \volavenf) -1.08 \volavenf ,
\label{eq:sigfit}
\end{eqnarray}
where $M = M_{10} \times 10^{10}\Msun$.  Note, however, that this fit only applies to halos at $z=9$; the dependence on mass will no doubt change with redshift.  It is possible, however, that the dependence on $\volavenf$ is more robust, because the ionization pattern is nearly independent of the timing of reionization \citep{FMH05, McQuinn07}.

\section{z=9 LAE Luminosity Functions}
\label{sec:lf}

With our optical depth distributions in-hand, we can proceed to generate $z=9$ LAE luminosity functions.  To do this, we make the standard simplifying assumption (e.g. \citealt{FZH06, DWH06, SLE07, McQuinn07LAE}) of a deterministic, linear mapping of halo mass to \lya\ luminosity, $M \propto L$.  As mentioned above, this assumption is often justified by the narrow mass range probed by existing instruments.  Using this ansatz, the observed \lya\ luminosity of a LAE, $L_{\rm obs} = L e^{-\taudamp}$ (where the ``intrinsic'' \lya\ luminosity, $L$, includes resonant attenuation), can be written in terms of the halo mass: $M_{\rm obs} = M e^{-\taudamp}$, where $M_{\rm obs}$ is the apparent mass of the halo under this mapping, after attenuation by the IGM.  Given the minimum observable luminosity, $L(\Mmin)$ (where $M_{\rm min}$ is the mass that would produce a luminosity $L$ without any IGM absorption), one could then detect halos with $\taudamp < - \ln(\Mmin / M)$.  In order to maintain generality, we postpone the discussion of the $L(M)$ mapping until \S \ref{sec:obs}, where we compare the luminosity functions to observations.

Hence, the cumulative number density of observable halos can be written as:
\begin{eqnarray}
\label{eq:lf}
n(>\Mmin, \avenf) &=& \int_{\Mmin}^\infty dM \frac{dn(>M)}{dM}\\
\nonumber &\times& \int_0^{\ln (M/\Mmin)} d\taudamp \frac{\partial p(>\taudamp, M, \avenf)}{\partial \taudamp}  ~ ,
\end{eqnarray}
\noindent where $dn(>M)/dM$ is the number density of halos per unit mass, and the second integral is the fraction of halos of mass $M$ which can be detected above the threshold $\Mmin$.

\begin{figure}
\vspace{+0\baselineskip}
\myputfigure{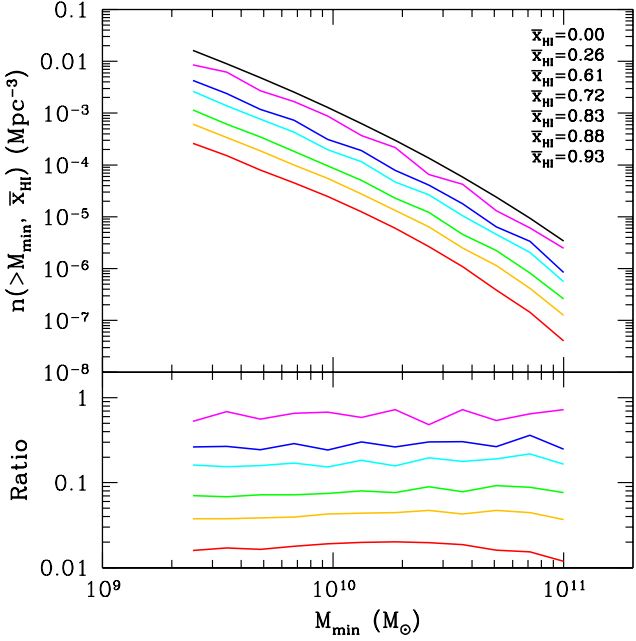}{3.3}{0.5}{.}{0.}  \caption{ 
Luminosity functions at $z=9$.  {\it Top:} Number density of objects brighter than some observed luminosity $L_{\rm obs}(\Mmin)$, i.e. with $M$ exp[$-\tau_D] > \Mmin$. 
Curves correspond to $\avenf \approx$ 0, 0.26, 0.61, 0.72, 0.83, 0.88, 0.93 ({\it top to bottom}).
{\it Bottom:} Ratio of luminosity functions: $n(>\Mmin, \avenf)$/$n(>\Mmin, 0)$.  
 Curves correspond to $\avenf$ = 0.26, 0.61, 0.72, 0.83, 0.88, 0.93 ({\it top to bottom}).
\label{fig:lf}}
\vspace{-1\baselineskip}
\end{figure}

``Luminosity" functions generated with this procedure are shown in the top panel of Figure \ref{fig:lf}. The curves correspond to $\avenf \approx$ 0, 0.26, 0.61, 0.72, 0.83, 0.88, 0.93 ({\it top to bottom}).
The bottom panel shows the ratio of the apparent and intrinsic luminosity functions: $n(>\Mmin, \avenf)$/$n(>\Mmin, 0)$.  Curves correspond to $\avenf$ = 0.26, 0.61, 0.72, 0.83, 0.88, 0.93 ({\it top to bottom}).

The luminosity functions in Fig. \ref{fig:lf} represent the first such predictions for $z=9$ which include the effect of inhomogeneous reionization.  As in analytic models and other simulations, the decline in the number density of observable LAE in a partially neutral universe when compared to a fully ionized universe (see bottom panel of Fig. \ref{fig:lf}) is nearly independent of halo mass \citep{FZH06, McQuinn07LAE}.\footnote{The slightly ``wavy'' features in some of the curves in Fig. \ref{fig:lf} result from the discrete halo masses returned by our halo filtering procedure.  The effect is more noticeable at small $\avenf$, when the $\taudamp$ distributions become weaker functions of $M$ (see Fig. \ref{fig:taud} and eq. \ref{eq:lf}).}  \citet{McQuinn07LAE} examined this question in detail and showed that the scale-independence is robust to other assumed mappings between halo mass and luminosity.  One explanation is that a lognormal distribution of optical depths, acting on a power-law intrinsic luminosity function, produces a nearly power-law apparent luminosity function.  The bright end, where the intrinsic luminosity function falls exponentially, would also fall exponentially except that the broad distribution of optical depths nearly erases the change in slope.  Our results are consistent with this explanation (see also \citealt{McQuinn07LAE}), especially because we have shown that the optical-depth distribution is broadest for massive objects.  The (nearly) scale-independent suppression appears to be a generic feature of reionization, requiring some other explanation for the decline in number density for bright LAEs observed by \citet{Kashikawa06}, and possibly that of \citet{ota07} as well.

\subsection{Existing Observational Constraints}
\label{sec:obs}

By taking advantage of the strong magnification provided by gravitational lensing through galaxy clusters, \citet{Stark07} found six candidate ($>5\sigma$) LAEs in the redshift range $z=$ 8.7--10.2, with all but one falling within $z = 9 \pm 0.35$.  The candidates have unlensed luminosity estimates ranging from 10$^{41.2}$ to 10$^{42.7}$ ergs s$^{-1}$. A search for additional emission lines expected if the candidates were low-$z$ interlopers has been completed and has not found evidence for a low-$z$ scenario for any of the six candidates.  If genuine, these detections can provide an invaluable first glimpse into the $z\sim9$ universe.

In order to compare our cumulative number densities in Fig. \ref{fig:lf} with observations, one needs a mapping of \lya\ luminosity, $L$, to halo mass, $M$.  Even if one assumes a deterministic, linear relation, there are still many unknowns.  The simplest version follows a well-trod path.  We begin by assuming that about 2/3 of the ionizing photons absorbed within the galaxy are converted into \lya\ photons \citep{Osterbrock89}.  One can then write the conversion as $L = 0.67 h \nu_{\alpha} (1-f_{\rm esc}) \dot{\rho}_\ast \epsilon_\gamma \Tgamma$, where $\nu_{\alpha}$ is the rest-frame \lya\ frequency, $f_{\rm esc}$ is the escape fraction of ionizing photons, $\dot{\rho}_\ast$ is the star formation rate (SFR), $\epsilon_\gamma$ is the ionizing photon efficiency per stellar mass, and $\Tgamma$ is the fraction of \lya\ photons which escape from the galaxy without getting {\it resonantly} absorbed.\footnote{As we have seen, our simulations model the damping wing component of the IGM absorption, $\taudamp$.  Hence, in order to compare with our simulated number densities, we only need worry about the fraction of \lya\ photons that are {\it resonantly} absorbed. Defined as such, $\Tgamma$ ranges approximately from 0.5 to 1.}  Assuming that galaxies steadily convert a fraction, $f_\ast$, of their gas into stars over some mean time-scale, $t_\ast$, and that $f_{\rm esc} \ll 1$, one can write the above relation as:
\begin{eqnarray}
\label{eq:MtoL}
L &=& 0.67 h \nu_{\alpha} f_\ast \frac{\Omega_b}{\Omega_{\rm M}} M \frac{1}{t_\ast} \epsilon_\gamma \Tgamma\\
\nonumber &=& 1.88 \times 10^{-12} {\rm erg} ~ \left( \frac{\epsilon_\gamma f_\ast \Tgamma}{t_\ast} \right) M ~ .
\end{eqnarray}
The free parameters in equation (\ref{eq:MtoL}) are all almost unconstrained at high redshifts.  Hence, we are interested in exploring a wide range of possibilities.  From an astrophysical standpoint, it is much easier to use observed number densities to set robust {\it upper limits} on $\avenf$ than it is to set robust lower limits, since in the theoretical model one at least has the hard upper limit of using all available gas in every galaxy above the detection threshold.  Setting conservative upper limits on $\avenf$ translates to maximizing $(\epsilon_\gamma f_\ast \Tgamma/t_\ast)$ in the $L\leftrightarrow M$ mapping.  Keeping this in mind, we apply four different $L\leftrightarrow M$ choices to the \citet{Stark07} sample, with values of $\epsilon_\gamma$ all taken from \citet{Schaerer03} assuming metallicities of $Z \sim 0.04$ $Z_\ast$ for Pop II and $Z \sim 10^{-7}$ $Z_\ast$ for Pop III star formation:

(i) $z\sim6$ parameters with Pop II stars.  Here we use values which have been shown to fit $z=5.7$ LAE luminosity functions fairly well \citep{DWH06, SLE07, McQuinn07LAE}. We set $f_\ast \Tgamma=0.1$, $t_\ast=$ 2/3 of the Hubble time $=1.87\times10^{16}$ s, and $\epsilon_\gamma=6.3\times10^{60}$ ionizing photons $\Msun^{-1}$  as might be expected from a Pop II IMF. This translates to the relation $L= 6.3\times10^{31} {\rm erg ~ s^{-1}}(M/\Msun)$, and to roughly 0.24 ($f_{\rm esc}$/0.02) ionizing photons per H atom.\footnote{We estimate the number of ionizing photons per H atom by $(\epsilon_\ast/n_{\rm H}) f_\ast f_{\rm esc} (\Omega_b/\Omega_{\rm M}) \int M dn(>M)/dM~ dM$, where $n_{\rm H}$ is the number density of hydrogen atoms, the integration extends over all halos at $z=9$ above the cooling threshold $\Tvir=10^4$ K, and we set $\Tgamma=1$ in order to be conservative.}

(ii) $z\sim6$ parameters with Pop III stars.  Here we again take values of $f_\ast$, $\Tgamma$, and $t_\ast$ which fit $z=5.7$ LAE luminosity functions but use a Pop III IMF for $\epsilon_\gamma$ at $z=9$.  \citet{SLE07} show that a similar model can fit the \citet{Stark07} constraints moderately well. Specifically, we set $f_\ast \Tgamma=0.1$, $t_\ast=$ 2/3 of the Hubble time $=1.87\times10^{16}$ s, and $\epsilon_\gamma=8.1\times10^{61}$ ionizing photons $\Msun^{-1}$  as might be expected from a Pop III IMF.  This translates to the relation $L= 8.2\times10^{32} {\rm erg ~ s^{-1}}(M/\Msun)$, and to 3.1 ($f_{\rm esc}$/0.02) ionizing photons per H atom.

(iii) ``maximally'' conservative with Pop II stars. Here we push the limits of a Pop II model by setting $f_\ast \Tgamma=1$, $t_\ast=$ the dynamical time $=4.6\times10^{15}$ s, and $\epsilon_\gamma=6.3\times10^{60}$ ionizing photons $\Msun^{-1}$ as might be expected from a Pop II IMF.  
In other words, we assume that every baryon inside every halo is converted into stars in one dynamical time (with this star formation episode synchronized across all halos) and every \lya\ photon escapes the environs of the galaxy.  This translates to the relation $L= 2.6\times10^{33} {\rm erg ~ s^{-1}}(M/\Msun)$, and to 2.4 ($f_{\rm esc}$/0.02) ionizing photons per H atom.

(iv) ``maximally'' conservative with Pop III stars. Here we completely push the limits by setting $f_\ast \Tgamma=1$, $t_\ast=$ the dynamical time $=4.6\times10^{15}$ s, {\it and} $\epsilon_\gamma=8.1\times10^{61}$ ionizing photons $\Msun^{-1}$ as might be expected from a Pop III IMF. This translates to the relation $L= 3.3\times10^{34} {\rm erg ~ s^{-1}}(M/\Msun)$, and to 31 ($f_{\rm esc}$/0.02) ionizing photons per H atom.\footnote{We will see that this extreme mapping implies that the \citet{Stark07} galaxies are below the $\Tvir=10^4$ K atomic cooling threshold.  If we instead allow stars to form down to the H$_2$ cooling threshold, $\Tvir\sim300$ K,  this mapping would imply a whopping 140 ($f_{\rm esc}$/0.02) ionizing photons per H atom!}

The relations implied by (ii) or (iii) are probably the most reasonably conservative estimates, 
as any stronger sources would have produced many more photons than are required to reionize the IGM.  Relations with more efficient star formation [such as the extreme (iv)] would result in an early reionization, contrary to the evidence suggested by {\it WMAP} \citep{Page06} and the SDSS quasar spectra \citep{Fan06, MH04, MH07}.

\begin{figure}
\vspace{+0\baselineskip}
\myputfigure{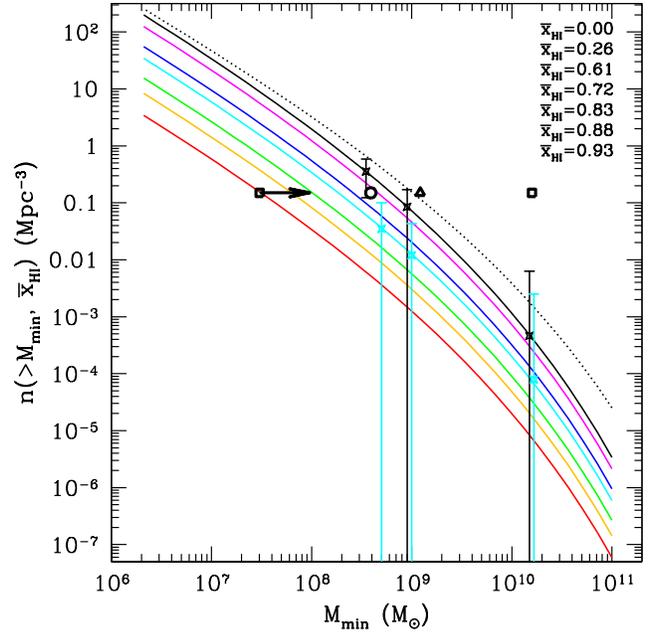}{3.3}{0.5}{.}{0.}  \caption{ 
Luminosity functions at $z=9$. Points correspond to number densities derived from the six $z\sim9$ candidates in the \citet{Stark07} survey, mapped using the $L\leftrightarrow M$ relations in the text, with (i), (ii), (iii), (iv) mappings shown right to left in the figure.
Solid curves correspond to $\avenf \approx$ 0, 0.26, 0.61, 0.72, 0.83, 0.88, 0.93 ({\it top to bottom}).  The dotted curve corresponds to $\avenf \approx 0$, but with $\sigma_8=0.86$.
Total error bars (Poisson and cosmic variance), generated with the Monte-Carlo procedure described in text, are shown for $\avenf \approx$ 0 and 0.72, at several choices of $\Mmin$.
\label{fig:lf_stark}}
\vspace{-1\baselineskip}
\end{figure}

Number densities derived from the $z\sim9$ candidates in the \citet{Stark07} survey, transformed to mass units using the $L\leftrightarrow M$ relations above are shown in Figure \ref{fig:lf_stark}, with the (i), (ii), (iii), (iv) mappings shown right to left in the figure \footnote{Of the three luminosity bins presented in \citet{Stark07}, we only show the most tightly constrained one at $L=10^{42}$ erg s$^{-1}$  This is the only number density with less than 100\% Poisson errors.}.  Note that extending the $\avenf>0$ luminosity functions from our model to the low scales required by the most conservative $L\leftrightarrow M$ relation is computationally prohibitive.  Luckily, the suppression of the luminosity function is almost independent of $M$ (see the bottom panel of Fig. \ref{fig:lf}), allowing us to extend the $\avenf>0$ estimates by using the mean suppression ratios obtained over the mass range $10^9$ -- $10^{11}$ $\Msun$. Solid curves thusly generated correspond to $\avenf \approx$ 0, 0.26, 0.61, 0.72, 0.83, 0.88, 0.93 ({\it top to bottom}).  

To estimate the size of the total (Poisson and cosmic variance) uncertainties, we have performed Monte-Carlo simulations modeling the \citet{Stark07} observations, which used the lensing signature of 9 clusters to probe a combined volume of 13.5 Mpc$^3$, given $L>10^{42}$ erg s$^{-1}$.  Specifically, we tile our simulation box with cells whose volume is equal to the mean volume probed by each cluster, 1.5 Mpc$^3$.\footnote{Note that the estimated geometry of this lensed long-slit spectroscopic survey can be approximated with a long parallelepiped with a 0.02 Mpc$^2$ FOV and a LOS distance of 500 Mpc per cluster \citep{SLE07}.  Since our box is 250 Mpc on a side and our halo field resolution scale is $\sim$0.17 Mpc, we must content ourselves with a flatter cell of the same volume with which to tile our box: a 0.03 Mpc$^2$ FoV with a LOS distance of 50 Mpc.}
We then repeatedly randomly select 9 such sample volumes in our simulation box, keeping track of the total number of LAEs in each group of 9 sample 
volumes.\footnote{Note that this tiling procedure does not precisely mimic the survey, because we should ideally select volumes without tiling beforehand.  Obtaining accurate statistics by randomly sampling volumes of our simulation box would be computationally prohibitive at small source number densities.}  The total error bars thus generated for $\avenf \approx$ 0 and 0.72, at several choices of $\Mmin$ are now shown in Fig. \ref{fig:lf_stark}.

\begin{figure*}
\begin{center}
\resizebox{8cm}{!}{\includegraphics{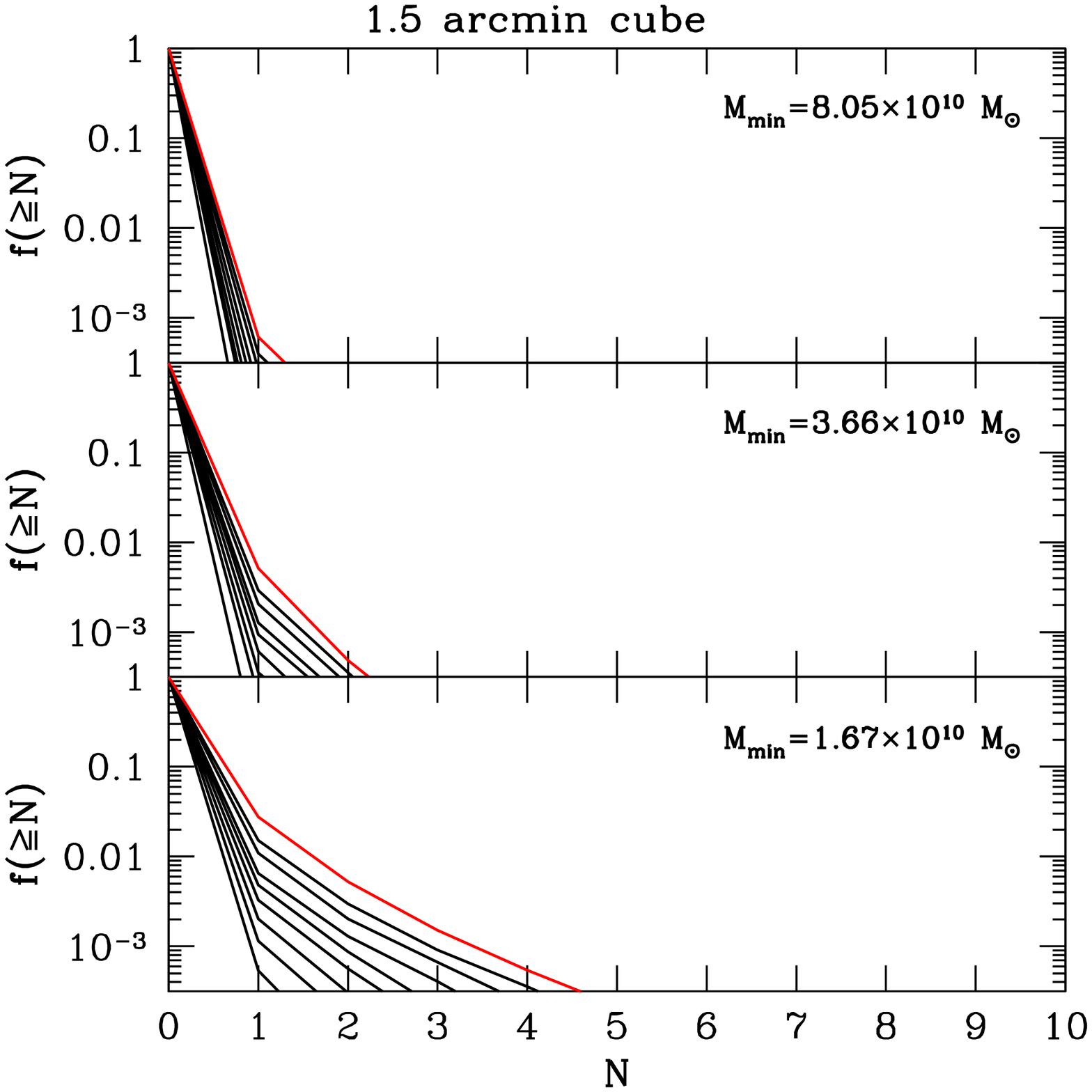}}
\hspace{0.13cm}
\resizebox{8cm}{!}{\includegraphics{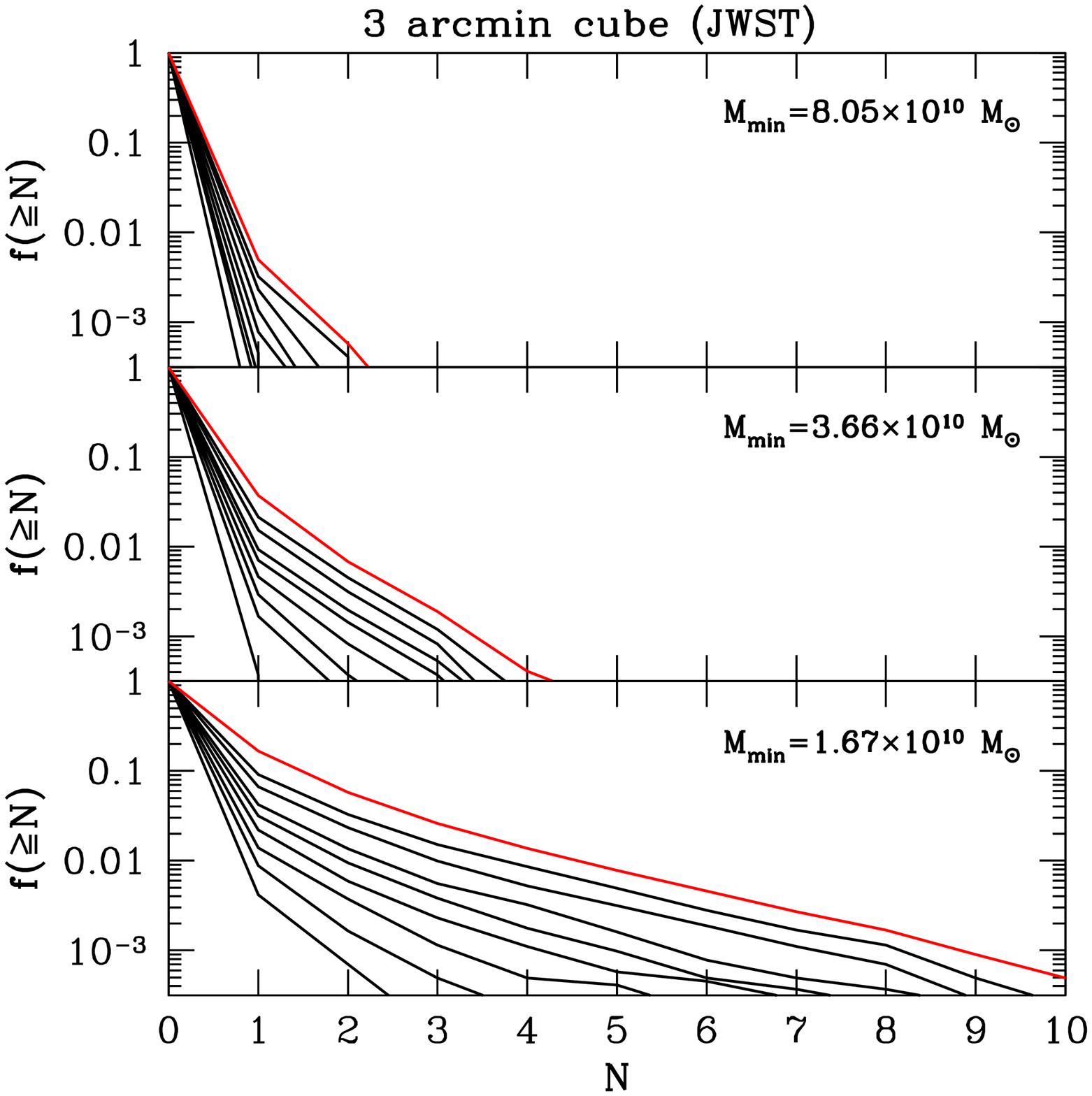}}
\end{center}
\caption{
Fraction of mock-survey fields containing $\ge N$ number of LAE.  Curves correspond to $\avenf \approx$ 0, 0.46, 0.56, 0.67, 0.72, 0.77, 0.83, 0.88, 0.93 ({\it top to bottom}), assuming $\Mmin =$ $8.06\times10^{10}$, $3.68\times10^{10}$, and $1.67\times10^{10}$ $\Msun$ in the sub-panels ({\it top to bottom}).  Panels assume FoVs of: $1.5'\times1.5'$ ({\it left panel}) and $3'\times3'$ ({\it right panel}).
\label{fig:incellpdf}
}
\end{figure*}

Assuming that the \citet{Stark07} candidates are genuine and using the maximally conservative relation in (iv), $\avenf(z\sim9)$ cannot be constrained:  such a scenario would permit the IGM to be almost entirely neutral.  However, it is very unlikely that {\it all} the factors in eq. (\ref{eq:MtoL}) conspire to provide such efficient star formation and \lya\ photon transmission
(and even if they did, reionization would likely have ended long before).  
The more reasonably conservative models, (ii) and (iii), prefer $\avenf \lsim 0.2$ and $\avenf \lsim 0.7$, respectively.  The $L\leftrightarrow M$ relation from (i), which has been shown to provide a decent fit to $z=5.7$ LAE luminosity functions, requires quite massive halos and drastically overestimates the observed $z\sim9$ abundances (overestimating the mass function by a factor of $\sim$400 at the required masses).  It is therefore inconsistent with even a completely ionized universe. Note however that since our estimates are based on the limited volume of our 250 Mpc$^3$ simulation box, we might be slightly underestimating the cosmic variance contribution to these error bars.

One possible remedy is if our background cosmology is wrong.  The dotted curve shows the mass function if $\avenf \approx 0$, but with $\sigma_8=0.86$, as the combined three--year {\it WMAP} and Lyman--$\alpha$ forest data prefer \citep{Lewis06}.  This higher value of $\sigma_8=0.86$ increases the abundance by a factor of several at the high-mass end, but it has only a modest effect in the range of interest.  The observed abundances in our model (i) are still inconsistent with a $\avenf \approx 0$ universe, overestimating the mass function by a factor of $\sim$100.  

On a more fundamental level, the \citet{Stark07} LAE sample, if genuine, \emph{requires} substantial star formation in halos with $M \lsim 10^9 \Msun$.  This is only a factor $\sim10$ greater than the atomic cooling threshold at $z\sim9$, and these objects would correspond to some of the smallest galaxies observed at \emph{any} redshift.  They must also be fundamentally different from sources at $z=5.7$, with either much higher star formation rates or stars that are much more efficient at producing ionizing photons.  Unless the escape fraction of UV photons is extremely small, any scenario that is consistent with the observations would produce several ionizing photons per IGM baryon, so we would expect reionization to have been well underway (if not complete) by $z=9$.  Needless to say, such a scenario would be difficult to reconcile with reionization at $z \sim 6$--$7$.

To this point, we have assumed that \emph{all} galaxies above $M_{\rm min}$ are LAEs.  However, at moderate redshifts, we know that a majority of Lyman-break galaxies (LBGs) have weak or absent \lya emission lines (e.g. \citealt{Shapley03}), though there is some evidence of an increase in the fraction of LAEs among LBGs at higher redshifts \citep{Dawson04, Hu04, Shimasaku06}. Incorporating this into Figure~\ref{fig:lf_stark} would be equivalent to shifting the curves downward by an amount equal to the fraction of all galaxies that are LAEs.  Thus this only strengthens our arguments, because it \emph{widens} the disparity between the observed number density of sources and the theoretical curves.

\section{Counts-in-cells Statistics}
\label{sec:incell}

LAE abundances and luminosity functions have certainly proven to be very useful in studying the $z\sim6$ universe.  However, their interpretation is inevitably controversial, because knowledge of the $M\leftrightarrow L$ mapping is required for a meaningful estimate of $\avenf$.  We have already seen how difficult it is to place meaningful constraints with this method.

It is therefore worth considering other, more robust, signatures.  Reionization modulates the observed LAE field, increasing the clustering of sources \citep{FHZ04, FZH06, McQuinn07LAE}.  Galaxies inside large HII bubbles are more likely to be seen than those inside smaller bubbles.  Thus, during reionization, we expect a field of view (FoV) that contains a LAE to have a {\it higher} probability of containing {\it another} LAE, than would be the case after reionization.  This reionization-induced clustering is illustrated in Figure~\ref{fig:maps} and can be quantified in any number of ways.  To date, studies have focused on the linear bias and power spectrum \citep{FZH06, McQuinn07LAE}.  However, the modulation is non-gaussian, so other statistics may be as powerful.  

Here we explore simple, statistical estimates of ``counts-in-cells'' that can be easily applied to future surveys.  These simple number counts essentially represent an ``integrated'' measurement of the clustering and hence can be easier to generate with a limited observational sample than other statistical indicators such as the power spectrum/correlation function, which try to measure the detailed scale dependence.  They also place less stringent requirements on the survey strategy than, for example, the power spectrum, because they are easier to interpret with non-uniform survey coverage (for example, following up bright sources to search for fainter neighbors; see below).
They are therefore most likely to be powerful early in reionization or at extremely high redshifts, when sources are rare.  

\begin{figure}
\vspace{+0\baselineskip}
\myputfigure{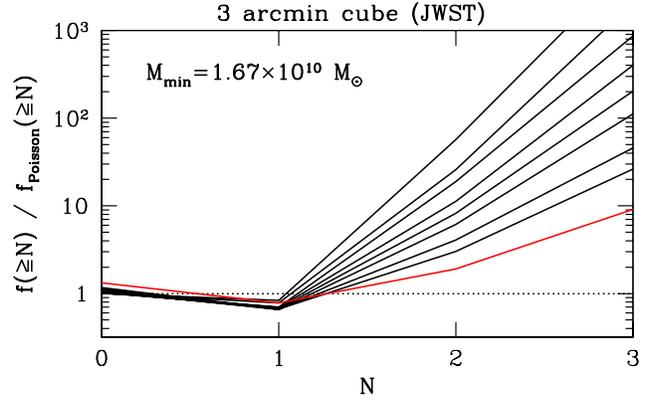}{3.3}{0.5}{.}{0.}  \caption{ 
Ratio of the PDF curves from the bottom right panel in Fig. \ref{fig:incellpdf} (in reverse order, {\it bottom to top}) to those expected from a pure Poisson distribution.
\label{fig:poisson}}
\vspace{-1\baselineskip}
\end{figure}

To study this reionization-induced clustering, we perform a mock-survey in our simulation box.  We tile our simulation box and tabulate ``in-cell'' number counts, given $\avenf$ and $\Mmin$.  For simplicity, we assume the survey FoVs are cubical volumes, but our results can easily be extended to more detailed survey specifications, once they are available.  Note also that cell sizes are fairly arbitrary, as surveys can be broken down into small cells for analysis, and the optimal choice will depend on the survey depth and volume.
In Figure \ref{fig:incellpdf}, we present the fraction of our mock-survey fields containing $\ge N$ LAEs.  The panels are constructed assuming FoVs of  $1.5'\times1.5'$ ({\it left panel}) and $3'\times3'$ ({\it right panel}), corresponding to comoving sizes of 4.23$^3$ and 8.45$^3$ Mpc$^3$, respectively. The latter (in angular coordinates) is the FoV of the Near-Infrared Spectrograph (NIRSpec) on {\it JWST} \citep{Gardner06}; other future IR instruments subtend comparable areas (for example, the Infrared Multi-Object Spectrograph (IRMOS) on the proposed TMT has a $5' \times 5'$ FoV\footnote{http://www.tmt.org}).  Curves correspond to $\avenf \approx$ 0, 0.46, 0.56, 0.67, 0.72, 0.77, 0.83, 0.88, 0.93 ({\it top to bottom}), assuming $\Mmin =$ $8.06\times10^{10}$, $3.68\times10^{10}$, and $1.67\times10^{10}$ $\Msun$ in the sub-panels ({\it top to bottom}).
  In Figure \ref{fig:poisson}, we show the ratio of the PDF curves from the bottom right panel in Fig. \ref{fig:incellpdf} to those expected from a pure Poisson distribution.
Note the relatively small change in the probabilities during the later stages of reionization, from $\avenf=0.5$ to $\avenf \approx 0$, compared to the rapid evolution during the initial stages.  This is, of course, primarily because the strong damping-wing absorption causes most of the sources to fall below the survey detection threshold before $\avenf \sim 0.5$.

In other words, because the curves shown in Fig. \ref{fig:incellpdf} correspond to raw number counts, they include two distinct consequences of higher values of $\avenf$: (i) the enhanced clustering footprint of reionization; and (ii) the decrease in the absolute number of observed LAEs.  As is the case for luminosity functions, (ii) is difficult to disentangle from astrophysical uncertainties and hence is not an ideal probe of reionization.  The first offers more promise, as the enhanced clustering of LAEs from reionization could only be mimicked by a large shift in the masses of underlying halos hosting LAEs, as discussed below. Thus we wish to isolate the first effect.

\begin{figure*}
\begin{center}
\resizebox{8cm}{!}{\includegraphics{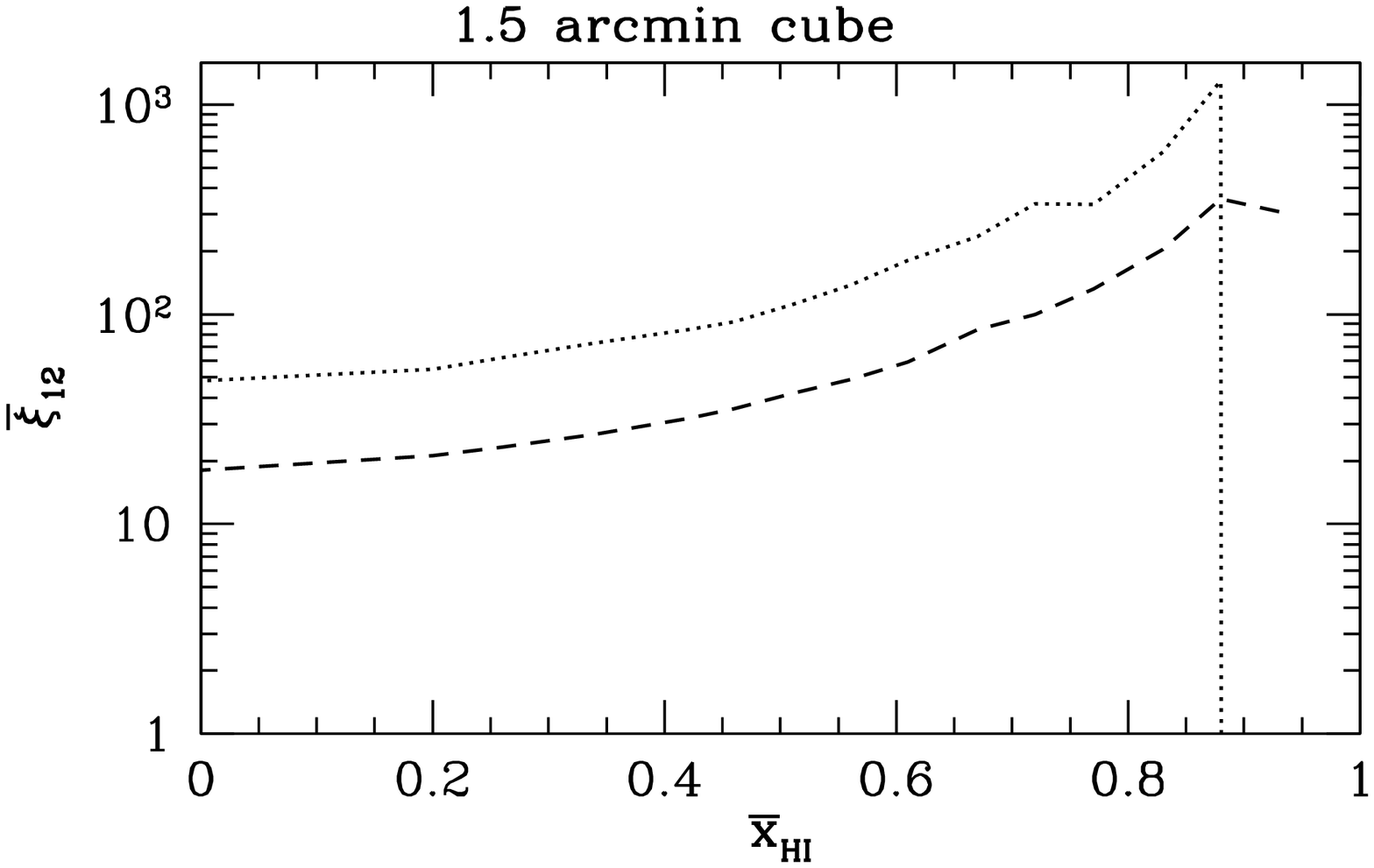}}
\hspace{0.13cm}
\resizebox{8cm}{!}{\includegraphics{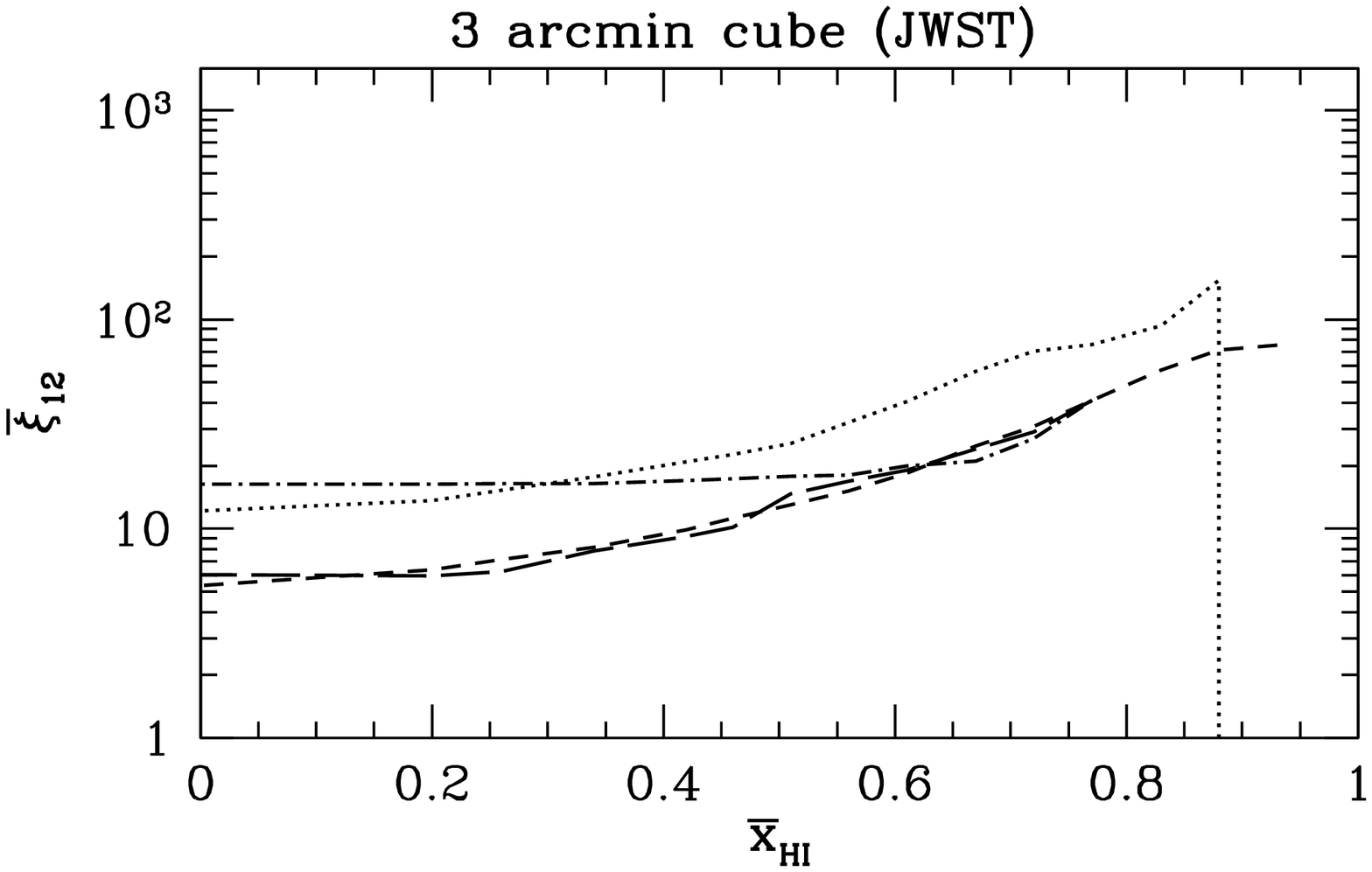}}
\end{center}
\caption{
The average two-point correlation function from eq. (\ref{eq:avexi}) as functions of $\avenf$.  Curves correspond to $3.68\times10^{10}$ $\Msun$ ({\it dotted}), and $1.67\times10^{10}$ $\Msun$ ({\it short-dashed}).  The long-dashed curve in the right panel shows the same quantity as the short-dashed curve, but in a scenario in which $\langle N \rangle$ is held constant at the number density found with $\avenf = 0.77$.  The dot-dashed curve in the right panel is generated assuming $\Mmin=1.67\times10^{10}$ $\Msun$, but by selecting only the most massive halos with number density fixed by the value at $\avenf = 0.77$.
\label{fig:sigma}
}
\end{figure*}

To this end, we note that the variance of the in-cell count distributions can be represented as (e.g. \citealt{Peebles80}):

\begin{equation}
\label{eq:varCIC}
\sigma^2_N = \langle N \rangle + \left( \frac{\langle N \rangle} {V} \right) ^2 \int_{V} \int_{V} dV_1 dV_2 ~ \xi_{12} ~ ,
\end{equation}

\noindent where $\langle N \rangle$ is the mean number of LAEs in a cell (i.e. field), and the integrals over the two-point correlation function, $\xi_{12}$, are performed over the cell volume, $V$.  Keeping this in mind, in Figure \ref{fig:sigma}, we plot:

\begin{equation}
\label{eq:avexi}
\avexi \equiv ~ ( \sigma^2_N - \langle N \rangle ) ~ /~ \langle N \rangle^2 ~ ,
\end{equation}

\noindent essentially the average value of the correlation function over a cell's volume.  Curves correspond to $\Mmin=$ $3.68\times10^{10}$ $\Msun$ ({\it dotted}), and $1.67\times10^{10}$ $\Msun$ ({\it short-dashed}).

 If sources are uncorrelated (i.e. Poisson distributed),  $\sigma^2_N = \langle N \rangle$, and $\avexi = 0$.  
 The biased nature of structure formation manifests as a constant, positive $\avexi$ (because we hold redshift, and hence the halo population, fixed), and reionization manifests as an increase in $\avexi$. Shallower surveys with larger $\Mmin$ have a larger $\avexi$, because these rarer objects are more highly clustered intrinsically.  Interestingly, the curves are fairly flat for small neutral fractions; this is because nearly all of the objects are visible, so the observed clustering nearly matches its intrinsic value.  Reionization induces a sharp rise in $\avexi$ at $\avenf \sim 0.5$, with the location of the rise being a very weak function of the cell size and $\Mmin$.  This increase is a result of the ``small-scale" clustering enhancement described by \citet{FZH06}, which occurs because only the sources inside rare large bubbles remain visible early in reionization, but at the same time a large fraction of sources inside such bubbles are visible.  The enhancement is nearly independent of the underlying halo mass.

The long-dashed curve in the right panel of Fig. \ref{fig:sigma} shows the same quantity as the short-dashed curve, but in a scenario in which $\langle N \rangle$ is held constant regardless of the neutral fraction (at the number density found with $\avenf = 0.77$) by randomly selecting halos above the $\avenf$-dependent mass threshold.  The fact that the long and short dashed curves overlap illustrates explicitly that we have accurately removed the Poisson component of the fluctuations from our statistic.

The dot-dashed curve in the right panel of Fig. \ref{fig:sigma} is generated 
 by selecting only the most massive halos with number density fixed by the value at $\avenf = 0.77$ for $\Mmin=1.67\times10^{10}$ $\Msun$ (unlike the random selection performed for the long-dashed curve).  The curve is flat at small $\avenf$ since it corresponds to the same set of massive sources.  The curve then increases at $\avenf \gsim 0.6$, when ionized regions are small enough to make some of these highly-biased, massive sources fall below the the $\Mmin=1.67\times10^{10}$ $\Msun$ detection threshold. We see that the reionization-induced rise in $\avexi$ surpasses the intrinsic clustering of even the most massive sources with the same number density at $\avenf\gsim0.6$.  
Thus there is no question that reionization can be observed through clustering measurements, at least sufficiently early in the process.  Hence, counts-in-cells can provide a simple, robust probe of reionization.

Detecting the sharp rises evident in Fig. \ref{fig:sigma} could be a ``smoking-gun'' signature of reionization; however, at any particular redshift we have only one measurement and (as with the luminosity function) can only compare with measurements at different redshifts.\footnote{We remind the reader that our statistics are generated with the same intrinsic source field, i.e. at a fixed redshift $z=9$.  Since present-day simulations cannot accurately simulate the redshift evolution of $\avenf$, this is the cleanest way of extracting statistics on reionization, especially given that the ionization topology at fixed $\avenf$ is almost independent of redshift in this range \citep{McQuinn07}.  Unfortunately, the real Universe is uncooperative on this point, and thus observations of the different stages of reionization must necessarily be from different redshifts.}.
  The fact that the curves in Fig. \ref{fig:sigma} are strong functions of $\Mmin$ complicates the interpretation of any such detection, because any increase in $\avexi$ with redshift could be because of galaxy evolution alone; for example, even if $M_{\rm min}$ remains constant with redshift, the bias of the objects would still increase with redshift.  This degeneracy can be overcome if reionization progresses rapidly compared to the underlying structure, or if one correlates the LAE field with a Lyman-break galaxy (LBG) field, since the detectability of LBGs should not be affected by changes in $\avenf$ (see \citealt{McQuinn07LAE} for discussion of similar issues with reference to the power spectrum).

Unfortunately, such a process is always uncertain.  Instead we favor measuring the excess probability (over that in a completely ionized universe) that a cell will contain $N$ or more LAEs, {\it given that} it already contains at least one LAE:
\begin{equation}
\label{eq:delP}
\Delta P_{\avenf}(\ge N | \ge 1) = P_{\avenf}(\ge N | \ge 1) - P_{\avenf \approx 0}(\ge N | \ge 1) ~ ,
\end{equation}
\noindent where $P(\ge N | \ge 1)$ and $P_{\avenf \approx 0}(\ge N | \ge 1)$ are the probabilities that a cell containing at least one LAE will contain $N$ or more LAEs, in a partially ionized and a fully ionized universe, respectively, {\it normalized to have the same number density of LAEs}.  We normalize our $\avenf \approx 0$ field by randomly choosing LAEs above $\Mmin$, until we obtain the same number of LAEs as in the partially ionized box.  Note, however, that the proper normalization procedure is not well-defined so long as the mapping between mass and luminosity remains unknown.  However, we checked that selecting only the most massive halos, until we obtain the same number density, did not appreciably change the results for the $N=2$ curves.

The basic idea of this measure is to remove the overall normalization of the galaxy number density, whose evolution is uncertain, but to retain the enhanced probability of observing groups of sources inside the rare large bubbles.  
Also note that while the $\avexi$ statistic in eq. (\ref{eq:avexi}) only uses second-order correlations and can be obtained by integrating over the power spectrum, the counts-in-cells method takes advantage of higher-order correlations and the excess probability from eq. (\ref{eq:delP}) shows this explicitly for $N>2$.

\begin{figure*}
\begin{center}
\resizebox{8cm}{!}{\includegraphics{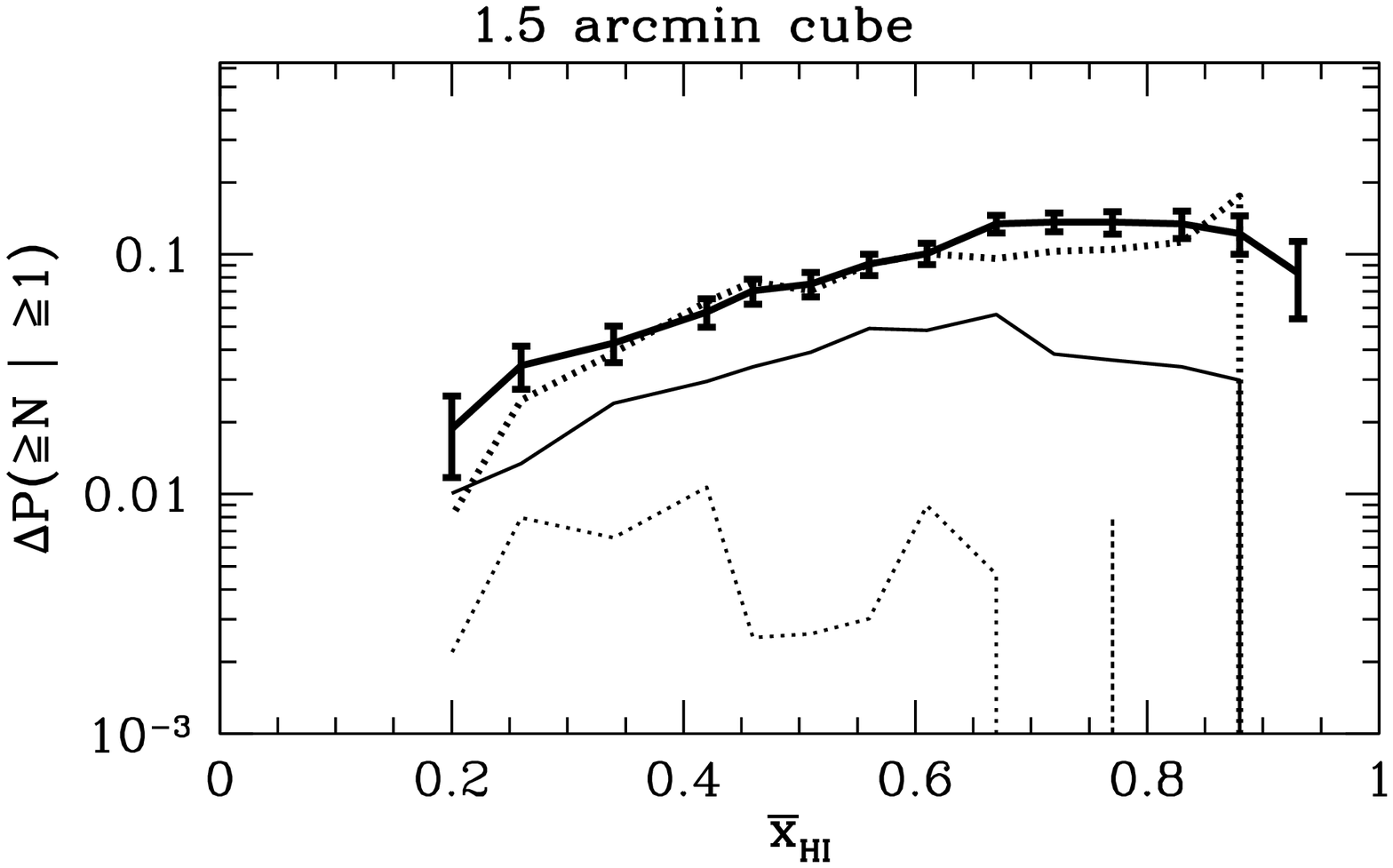}}
\hspace{0.13cm}
\resizebox{8cm}{!}{\includegraphics{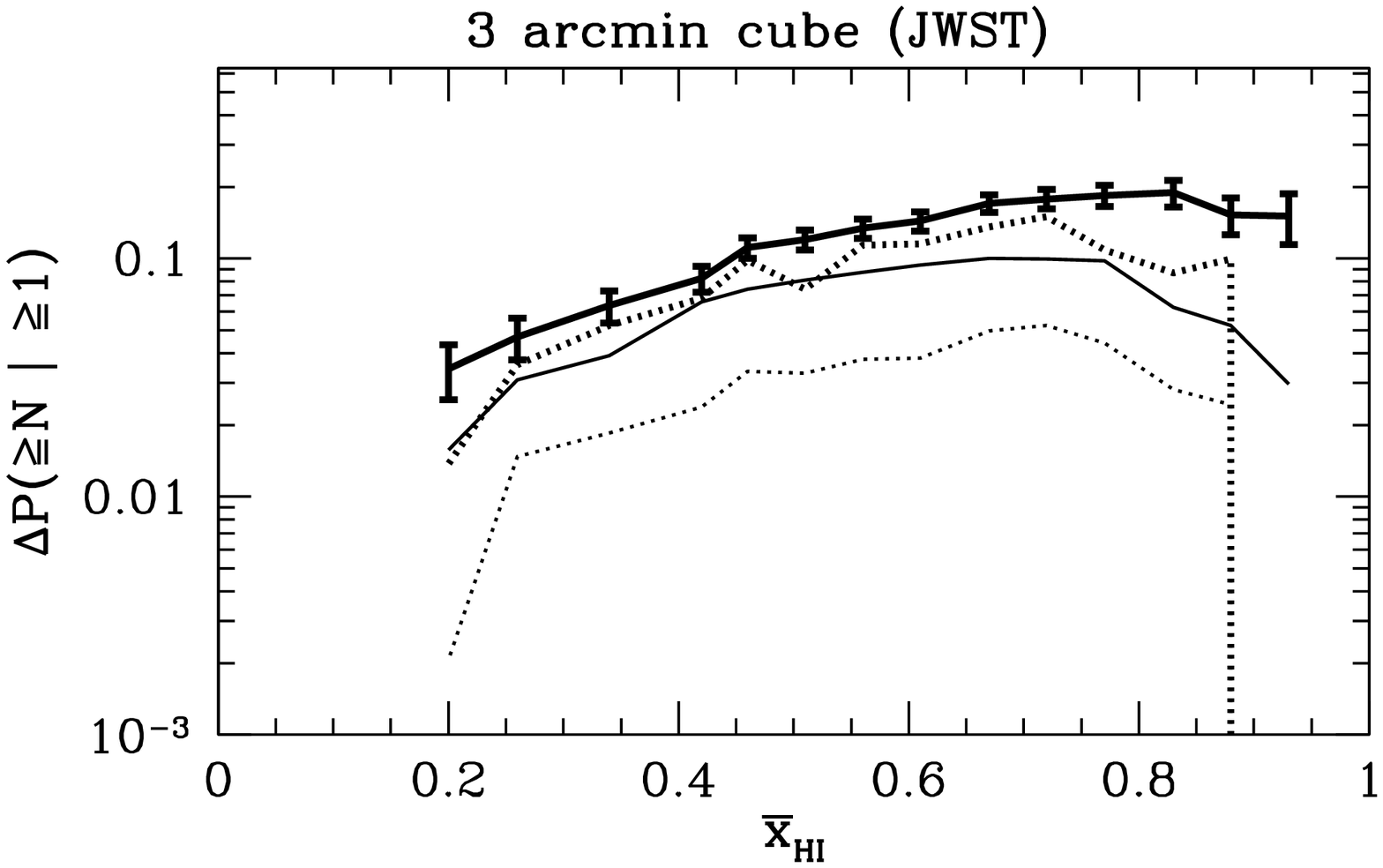}}
\end{center}
\caption{
Excess probability (over that in an ionized universe normalized to the same number density of LAEs) that a FoV containing at least one LAE will contain $N$ or more LAEs.   Panels correspond to FoV of $1.5'\times1.5'$ ({\it left}) and $3'\times3'$ ({\it right}). Solid and dotted curves assume $\Mmin= $ $1.67\times10^{10}$ and $3.68\times10^{10}$ $\Msun$, respectively. Thick and thin curves correspond to $N=$ 2 and 3, respectively. Error bars indicate the 1-$\sigma$ Poisson uncertainty on the $N=2$, $\Mmin= $ $1.67\times10^{10}$ $\Mmin$ curves.
\label{fig:delPvsnf}
}
\end{figure*}

In Figure \ref{fig:delPvsnf}, we plot the excess probability from eq. (\ref{eq:delP}), for  $1.5'\times1.5'$ ({\it left panel}) and $3'\times3'$ ({\it right panel}) cells. Solid (dotted) curves assume $\Mmin= $ $1.67\times10^{10}$ ($3.68\times10^{10}$) $\Msun$. Thick and thin curves correspond to $N=$ 2 and 3, respectively.

The enhanced clustering footprint of reionization can easily be seen from Fig. \ref{fig:delPvsnf}.  For $\avenf \gsim 0.5$ and $N=2$, $\Delta P \ga 10\%$, and it increases by a factor of $\gsim 6$ from $\avenf=$ 0.2 to 0.8.  Furthermore, the $N=2$ curves are much more sensitive to $\avenf$ than to $\Mmin$, suggesting that this statistic of the reionization-induced clustering \emph{cannot} be mimicked by a change in the mass threshold of the survey, or analogously by a shift in the underlying halo masses hosting LAEs.  Hence, the excess probability that a cell containing LAEs contains more than one LAE is a robust indicator of changes in $\avenf$.  The $N=3$ curves do fall significantly for the most massive objects (as do all the curves for $\avenf \approx 1$); this is simply because the sources are too rare (see Fig. \ref{fig:delPvsnf}).  The apparent drop-off at $\avenf \approx 0.9$ is an artifact of our finite box size.

We attempt to approximately remove the Poisson component of the clustering statistics from the excess probability by subtracting the second term in eq. (\ref{eq:delP}).  However, it is not immediately obvious that this statistic is completely independent of the LAE number density.  Hence it is intriguing to probe the robustness of the seeming overlap of the $N=2$ curves of different mass scales in Fig. \ref{fig:delPvsnf}.  To this end, we recreated the $\Mmin = 1.67 \times 10^{10}$ $\Msun$, $N=2$ curve from Fig. \ref{fig:delPvsnf} for $\avenf \leq 0.77$, by randomly choosing halos in order to keep the number density constant (i.e. equal to the number density at $\avenf = 0.77$, as we had for the long-dashed curve in the right panel of Fig. \ref{fig:sigma}).  We find that the excess probability remains fairly unchanged at $\avenf \gsim 0.5$, seemingly suggesting that our $\Delta P$ statistic is robust (not very sensitive to the LAE number density).  The excess probability falls off more rapidly at $\avenf \lsim 0.5$ (where the excess clustering signature is weaker) than in our Figure \ref{fig:delPvsnf}, though it is not clear if this trend is statistically significant since the error bars become large in this regime.  
Thus, our proposed $\Delta P$ statistic does appear to be relatively robust to the intrinsic number density, although the particularly close match in the $N=2$ curves of Figure~\ref{fig:delPvsnf} may be somewhat coincidental.

\begin{figure}
\vspace{+0\baselineskip}
\myputfigure{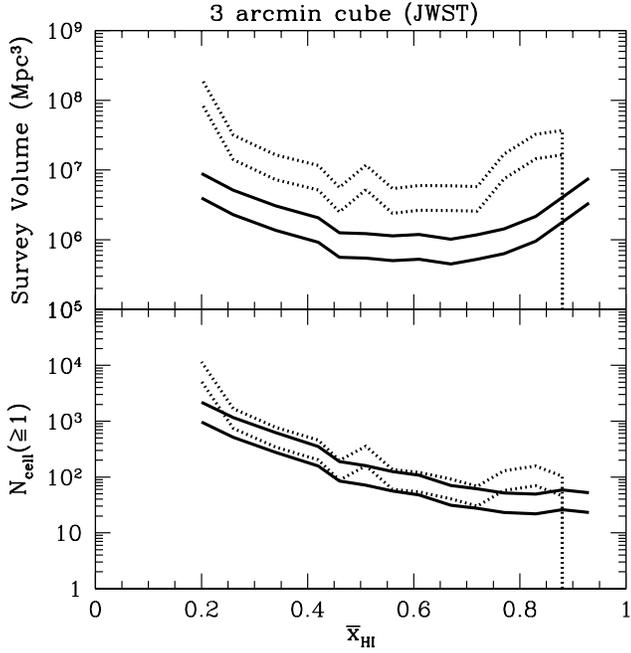}{3.3}{0.5}{.}{0.}  \caption{ 
Survey characteristics required to detect the excess clustering probability due to reionization (eq. \ref{eq:delP}) assuming $\Mmin =$ $1.67\times10^{10}$ $\Msun$ ({\it solid curves}) and $3.68\times10^{10}$ $\Msun$ ({\it dotted curves}), and $N=2$.  The top panel shows the survey volume required for 3-$\sigma$ ({\it top curve}) and 2-$\sigma$ ({\it bottom curve}) detections; the bottom panel shows the corresponding number of cells that actually contain galaxies.
\label{fig:Ntot}}
\vspace{-1\baselineskip}
\end{figure}

In Figure \ref{fig:Ntot}, we show the survey characteristics required to detect this excess probability, assuming $N=2$ and $3'$ cubical cells.  Specifically, we require $\Delta P - {\rm n}\sigma > 0$ for a n-$\sigma$ detection, where $\Delta P$ is our derived value from Fig. \ref{fig:delPvsnf} and $\sigma$ is the uncertainty on the measured value of $\Delta P$ in the given survey volume.  Solid and dotted curves correspond to $\Mmin= $ $1.67\times10^{10}$ and $3.68\times10^{10}$ $\Msun$, respectively.   In the bottom panel, we show the total number of cells containing galaxies that must be observed in order to detect the effect at the 3-$\sigma$ ({\it top curve}) and 2-$\sigma$ ({\it bottom curve}) level.  It is nearly independent of halo mass at higher neutral fractions, because this integrated clustering measure depends only weakly on the characteristics of the underlying halo population (see the discussion above).  Interestingly, a reasonably strong detection requires only several tens of galaxy detections, with the required source count decreasing as $\avenf$ increases.  This is because $\Delta P$ is large compared to the raw probability of detecting two neighboring galaxies and increases as the ionized regions get smaller.  This indicates the power of the counts-in-cell approach:  the actual number density of objects decreases by $\sim 5$ over this range, but this is compensated by the increased probability.  Many other approaches to clustering, such as the power spectrum, would lose sensitivity in this range because of the rarity of the sources.

The top panel shows the corresponding survey volumes that must be observed to detect this many galaxies.  (Of course, this increases rapidly with the mass threshold, even though $\Delta P$ does not, because the probability of having $N=1$ is much smaller for rarer sources.)  We can see that a survey volume of 6 (12) $\times 10^5$ Mpc$^{3}$ is required to detect the enhanced clustering at $\avenf\sim 0.5$--$0.8$, with a 2-$\sigma$ ($3$-$\sigma$) accuracy and $\Mmin= $ $1.67\times10^{10}$ $\Msun$ sensitivity.  If the survey depth is only one cell deep, this corresponds to a several square degree survey at least.  A deep, blind survey subtending such a region may be difficult in the foreseeable future, given the modest FoVs of forthcoming near-IR instruments.  However, the modest number of cells that actually contain sources suggests that a shallow but wide survey to identify extremely bright candidates, followed by deep followup with a large telescope to search for fainter neighbors, may be a viable strategy.  Moreover, if the \citet{Stark07} candidates are truly at $z=9$ and correspond to halos with $M \la 10^9 \Msun$, detecting large numbers of sources may be much easier than our conservative estimates suggest.  
This is one advantage of counts-in-cells over the power spectrum, which is more model-dependent in such unconventional survey strategies.

\begin{figure}
\vspace{+0\baselineskip}
\myputfigure{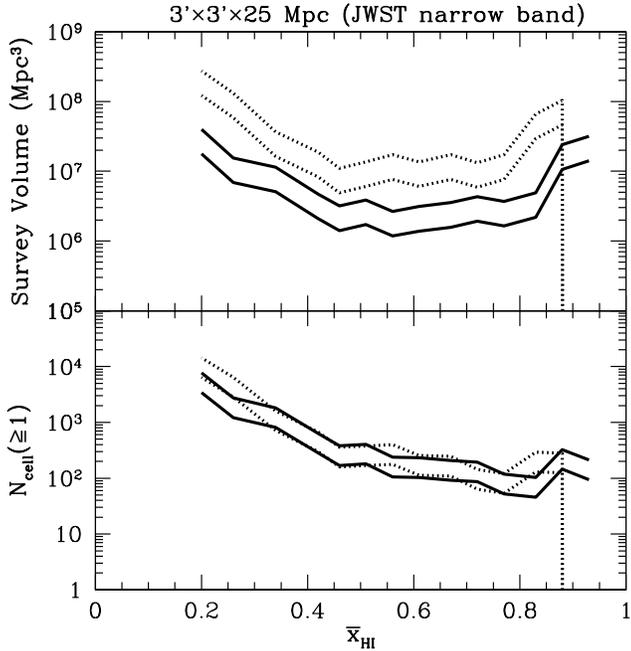}{3.3}{0.5}{.}{0.}  \caption{ 
Same as Fig. \ref{fig:Ntot}, but with a cell size typical of narrow band LAE surveys, with $R=\lambda/\Delta \lambda \sim 100$.
\label{fig:Ntot_narrowband}}
\vspace{-1\baselineskip}
\end{figure}

Throughout our discussion, we have used somewhat arbitrary cell sizes, as surveys can be broken down into small cells for analysis, and the optimal choice will depend on the characteristics of the particular survey.  However, in the absence of spectroscopic follow-up, the LAE redshift might only be localized to the width of a narrow band filter.  This sets a minimum cell size in the LOS direction.  
With this in mind, in Fig. \ref{fig:Ntot_narrowband}, we plot the same in-cell statistics as in Fig. \ref{fig:Ntot}, but extending the LOS axis of the cells to correspond to a narrow band filter with $R=\lambda/\Delta \lambda \sim 100$.  Comparing to Fig.~\ref{fig:Ntot}, the required number of cells and survey volume increases by less than a factor of $\sim$ 2.  Essentially, so long as the cells are not so long that random, distant neighbors overwhelm the reionization clustering, cells of any radial size can be used without a substantial increase in the survey requirements.

\begin{figure}
\vspace{+0\baselineskip}
\myputfigure{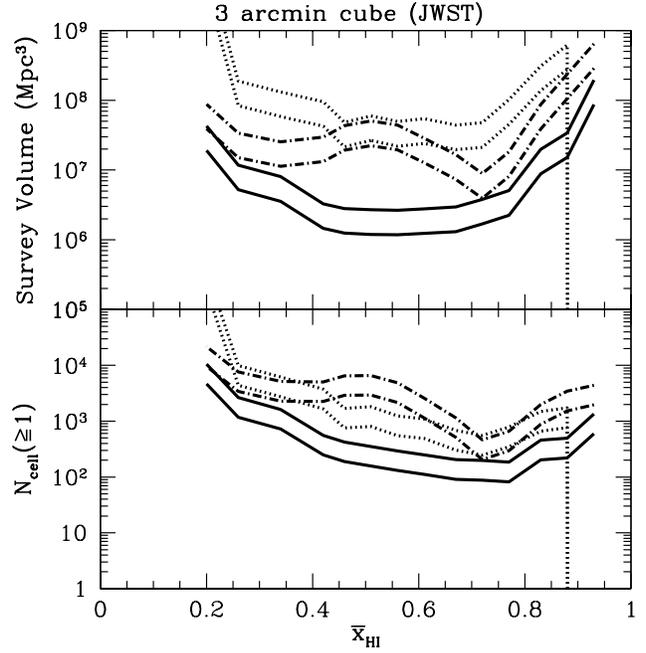}{3.3}{0.5}{.}{0.}  \caption{ 
Same as Fig. \ref{fig:Ntot}, but with $N=3$.  The dot-dashed curve is generated 
assuming $\Mmin=1.67\times10^{10}$ $\Msun$, but by selecting only the most massive halos, normalized to the same number density at each $\avenf$.
\label{fig:3point}}
\vspace{-1\baselineskip}
\end{figure}

As mentioned previously, counts-in-cells (and specifically our $\Delta P$ statistic) can make use of higher-order correlations. In Fig. \ref{fig:3point}, we plot the same statistics as in Fig. \ref{fig:Ntot}, but using clustering of the $N=3$ term.  Interestingly, at least for low-mass halos, measuring this term is no more difficult than measuring the two-point statistic.  These higher-order correlations have not yet been quantified in the context of reionization, so we also show (with the dot-dashed curve) the requirements (assuming $\Mmin=1.67\times10^{10}$ $\Msun$) to detect three-point clustering \emph{without} reionization, assuming that a survey probes all halos above a given mass threshold, normalized to the same number density.  (In other words, this is the requirement to detect three point clustering from just gravitational instability; for this nearly gaussian random field it contains little more information than the power spectrum itself.) The $N=3$ term is much weaker in this case and requires a considerably larger survey to measure.  Thus, detection of this three-point term may provide powerful support for any detection of reionization-induced clustering.

Note that we have ignored foreground contamination here.  This is likely a serious issue in any survey, but the details of foreground removal depend on the survey specifics.  Hence we defer it to future works, more focused on particular surveys.

As mentioned previously, stochasticity in the $L \leftrightarrow M$ mapping, and secondary effects on the detectability of the Ly$\alpha$ line, such as galactic winds and gas infall, can in principle affect the details of our counts in-cell estimates above.  We defer a thorough study of such effects to future works involving numerical simulations.

\section{Conclusions}
\label{sec:conc}

In this work, we investigate the use of LAEs in constraining reionization at $z\sim9$.  We generate $z=9$ halo, velocity and density fields in a 250 Mpc ``semi-numerical'' simulation box \citep{MF07}.  Our excursion-set approach allows us to resolve halos with masses $M\gsim2.2\times 10^8 \Msun$.  We construct ionization topologies corresponding to various values of $\avenf$.

As a first step, we generate damping wing \lya\ optical depth distributions for the halos in our simulations.  As expected, the more massive halos have narrower distributions with smaller mean absorption, though the distributions become weaker functions of halo mass as reionization progresses.
We show that these distributions are roughly lognormal but are broader and have a smaller mean than purely analytic predictions.

Using the $\taudamp$ distributions, we generate $z=9$ LAE luminosity functions for various values of $\avenf$.  Constraining $\avenf$ with luminosity functions is difficult due to the many uncertainties inherent in the host halo mass $\leftrightarrow$ \lya\ luminosity mapping.  
However, we show that applying a very conservative mapping to the number densities of the \citet{Stark07} sample yields $\avenf(z=9) \lsim 0.7$. More fundamentally, these LAE number densities, if genuine, \emph{require} substantial star formation in halos with $M \lsim 10^9 \Msun$, making them unique among the current sample of observed high-$z$ objects.

The topology of reionization increases the apparent clustering of the observed LAEs, aside from merely suppressing their number densities.  We investigate the detectability of this signature using ``counts-in-cell'' statistics, which are more robust than the power-spectrum at studying such non-gaussian fields and (as integrated measures) are more straightforward to interpret in the few source limit.  We find that the likelihood of observing {\it more than one} LAE among the subset of fields which contain LAEs is $\gsim$10\% greater in a universe with $\avenf \gsim 0.5$ than in an ionized universe with the same LAE number density.  We show that this effect can be detected at $z\sim9$ with just a few tens of $3'$ cubical cells containing galaxies, regardless of the underlying host halo mass.  

Counts-in-cells is only one approach to clustering, and it has advantages and disadvantages compared to the more common power spectrum (or correlation function) approach \citep{McQuinn07LAE}.  The latter accounts for the detailed scale dependence of the clustering enhancement (and so in principle provides information on the ionization field; \citealt{FZH06}) but does not include higher-order corrections to the clustering.  Moreover, robust power spectrum measurements require a large number of sources to be detected and are relatively unforgiving of ad hoc survey strategies, such as deep followup, which may be required when sources are rare.  The small FoVs of planned near-IR instruments therefore make such measurements difficult at high redshifts.

By contrast, our counts-in-cells approach offers very little information on the scale dependence of the ionization field.  However, it does include non-gaussianities, which become important early in reionization ($\avenf \ga 0.5$).  We explicitly showed, for the first time, that reionization induces non-gaussianities in the galaxy distribution that should be separable from structure formation, at least in the deep survey limit.  As a result, our signature becomes more powerful earlier in the reionization process, more than compensating for the declining apparent number density of LAEs.  It also places very few constraints on the survey geometry, because the fields need not be contiguous (and in fact, to compensate for cosmic variance, probably should not be).  An ideal strategy may be to identify particularly bright candidate objects with a wide, shallow survey.  Then, one can follow up these candidates with deeper integrations to confirm their identity and search for fainter neighbors.  Given that only a few tens of sources must be followed-up to provide interesting constraints, such a strategy would require relatively modest telescope resources.

\vskip+0.5in

We are grateful to D. Eisenstein for suggesting counts-in-cells as a probe of LAE clustering.  We thank D. Stark for providing his observed number densities of $z\sim9$ candidate LAEs.  We also thank M. 
McQuinn, D. Stark, and the anonymous referee for helpful comments on this manuscript.  This research was supported by NSF-AST-0607470.

\bibliographystyle{mn2e}
\bibliography{ms}

\end{document}